\documentclass[aps,prev, superscriptaddress, preprintnumbers,floatset,nofootinbib,twocolumn]{revtex4-1}
\pdfoutput=1
\usepackage[english]{babel}
\usepackage{bm}
\usepackage{times}
\usepackage{ulem}
\usepackage{hyperref}
\hypersetup{
    colorlinks=true,
    linkcolor=blue,
    filecolor=magenta,      
    urlcolor=blue,
    citecolor=blue,
    pdftitle={Sharelatex Example},
    pdfpagemode=FullScreen
    }
\usepackage{listings}
\usepackage{slashed}
\usepackage{color}
\usepackage{appendix}
\usepackage{mathtools}
\usepackage{epsfig}
\usepackage{dcolumn}
\usepackage{textcomp}
\usepackage{appendix} 
\usepackage{multirow}
\usepackage{graphicx}
\usepackage{soul}
\usepackage{subfigure}
\usepackage[compat=1.1.0]{tikz-feynman} 
\newcommand{\fig}[1]{Fig.~\ref{#1}} 
 
\newcommand{\eq}[1]{Eq.~(\ref{#1})} 

\begin{document} 

\title{The Hubble Rate Trouble: An Effective Field Theory of Dark Matter}

\author{Alvaro S. de Jesus}
\email{alvarosdj@ufrn.edu.br}
\affiliation{Departamento de F\'isica, Universidade Federal do Rio Grande do Norte, 59078-970, Natal, RN, Brasil}

\author{Nelson Pinto-Neto}
\email{nelsonpn@cbpf.br}
\affiliation{COSMO – Centro Brasileiro de Pesquisas F\'isicas, Xavier Sigaud, 150, Urca, 22290-180, Rio de Janeiro, Brasil}

\author{Farinaldo S. Queiroz}
\email{farinaldo.queiroz@ufrn.br}
\affiliation{Departamento de F\'isica, Universidade Federal do Rio Grande do Norte, 59078-970, Natal, RN, Brasil}
\affiliation{International Institute of Physics, Universidade Federal do Rio Grande do Norte,  59078-970, Natal, RN, Brasil}
\affiliation{Millennium Institute for Subatomic Physics at High-Energy Frontier (SAPHIR), Fernandez Concha 700, Santiago, Chile}

\author{Joseph Silk}
\email{silk@iap.fr}
\affiliation{Institut d’Astrophysique de Paris (UMR7095: CNRS \& UPMC- Sorbonne Universities), F-75014, Paris, France}
\affiliation{Department of Physics and Astronomy, The Johns Hopkins University Homewood Campus, Baltimore, MD 21218, USA}
\affiliation{BIPAC, Department of Physics, University of Oxford, Keble Road, Oxford OX1 3RH, UK}

\author{D\^eivid R. da Silva}
\email[Corresponding author: ]{deivid.rodrigo@academico.ufpb.br}
\affiliation{Departamento de Fisica, Universidade Federal da Paraiba, 58051-970, Jo\~ao Pessoa, PB, Brasil}

\begin{abstract}
The Hubble constant inferred from the 6-parameter fit to the CMB power spectrum conflicts with the value obtained from direct measurements via type Ia supernova and Cepheids observations. We write down effective operators involving spin-0, spin-1/2, and spin-1 dark matter that lead to the relativistic production of dark matter particles at early times, 
 and consequently lead to an increase in the number of relativistic degrees of freedom. This mechanism which is amenable to CMB, BBN, and structure formation observables can sufficiently raise the value of the Hubble constant derived from CMB and reconcile local and CMB probes of the Hubble constant. This mechanism alone increases $H_0$ up to  $70\, {\rm km s^{-1} Mpc^{-1}}$, and with the help of a Phantom-like cosmology, reach $H_0 \simeq 71-73\, {\rm km s^{-1} Mpc^{-1}}$. Lastly, we outline the region of parameter space which reproduces $H_0 \simeq 71-73\, {\rm km s^{-1} Mpc^{-1}}$ while obeying all relevant constraints.
 
\noindent

\end{abstract}


\maketitle
\flushbottom

\section{\label{In} Introduction}

The $\Lambda$CDM {cosmological model is grounded in the idea that the universe is nearly spatially flat, 
 with its structures arising from quantum vacuum fluctuations of cosmological perturbations from a highly homogeneous and isotropic primordial era. Currently, the universe is dominated by dark energy and cold dark matter \cite{piattella2018lecture,ref:hobson2006GR}. This simple description can nicely explain the abundance of light elements \cite{Cyburt:2015mya}, the CMB (Comic Microwave Background) power spectrum \cite{Planck:2018vyg}, the large scale structure as well as the ongoing accelerated expansion era \cite{DES:2020,Blanton:2017,Riess:1998,Perlmutter:1999}, among others. Nevertheless, an important discrepancy involving the Hubble constant surfaced. Considering the $\Lambda$CDM model and CMB power spectrum, Planck data favors $H_0 = 67.27 \pm 0.6$ km s$^{-1}$ Mpc$^{-1}$ \cite{Planck:2018vyg}. However,  adopting $\Lambda$CDM model, quasar time-delay cosmography leads to $H_0 = 71.9^{+ 2.4}_{-3.0}$ Km s$^{-1}$ Mpc$^{-1}$ \cite{H0LiCOW2016}. Parallax measurements of Cepheids provide $H_0 = 73.24 \pm 1.74$ km s$^{-1}$ Mpc$^{-1}$ \cite{ref:Bernal2016}. In summary, early measurements of the Hubble constant favor $H_0 < 69$ km s$^{-1}$ Mpc$^{-1}$, whereas local measurements yield $H_0 > 71$ Km s$^{-1}$ Mpc$^{-1}$ \cite{ref:Anchordoqui2021}. The incompatibility found in the Hubble constant is known as the Hubble tension, and its magnitude varies depending on the data set used. Table (\ref{tab:my_label}) presents some of these measurements showing a discrepancy between late and early universe data. Collectively speaking, it is clear that local measurements do not agree with CMB inferred values for $H_0$.

\begin{table}[h]
    \centering
    \begin{tabular}{|c|c|}
    \hline
    EARLY UNIVERSE&  Dataset\\
    \hline
    $H_0= 70.0 \pm 2.2 km s^{-1}Mpc^{-1}$  & WMAP9 \cite{Hinshaw_2013} \\
       $H_0=67.36 \pm 0.54 km s^{-1}Mpc^{-1}$ & CMB 2018 \cite{Planck:2018vyg}\\
    $H_0 = = 67.36 \pm 0.54 km s^{-1}Mpc^{-1}$  & SPT 2021 \cite{SPT-3G:2021eoc} \\
    $H_0=69.72 \pm 1.63 km s^{-1}Mpc^{-1}$  & ACT 2019 \cite{Wang:2019isw}\\
    $H_0= 67.9 \pm 1.1 km s^{-1}Mpc^{-1}$  & BOSS data \cite{Ivanov:2019pdj}\\
    $H_0= 69.6 \pm 1.8 km s^{-1}Mpc^{-1}$    & eBOSS Collab. \cite{eBOSS:2020yzd}\\
     \hline
    LATE UNIVERSE&  Dataset\\
     \hline
       $H_0= 73.8 \pm 2.1 km s^{-1}Mpc^{-1}$  & SN1a 2021 \cite{Cardona:2016ems} \\
       $H_0= 75.4 \pm 1.7 km s^{-1}Mpc^{-1}$ & Pantheon 2019 \cite{Camarena:2019moy}\\
    $H_0 = = 72.8 \pm 1.9 km s^{-1}Mpc^{-1}$  & Gaia 2020 \cite{Breuval:2020trd} \\
    $H_0=73.2 \pm 1.3 km s^{-1}Mpc^{-1}$  & Gaia and HST 2020 \cite{Riess:2020fzl}\\
    $H_0= 69.8 \pm 2.5 km s^{-1}Mpc^{-1}$  & Red Giants 2019 \cite{Freedman:2019jwv}\\
    \hline
    \end{tabular}
    \caption{ Early and late universe evaluations of the Hubble constant and their respective data sets.}
    \label{tab:my_label}
\end{table}

Studies raised systematic issues in the Planck analysis: One assumes the $\Lambda$CDM model to infer $H_0$ from the CMB data. One first example is the fact that Planck uses two different likelihood pipelines, Plik and CamSpec, which consider different sky masks and could, in principle, shift by $0.5 \sigma$ the $H_0$ constraints coming from the CMB. More importantly, the $A_{\rm lens}$ anomaly \cite{Calabrese:2008rt}, which is a nonphysical parameter equal to the unit if the gravitational lensing effects are the ones predicted by the $\Lambda$CDM, and null if there is no lensing at all. Planck collaboration sets $A_{\rm lens} > 1$ at two standard deviations. As this lensing anomaly has not been observed in the Planck trispectrum data, there is still unknown small systematic error in the CMB data, which could reduce the Hubble tension.

The late universe measurements of $H_0$ is direct, and come from measuring the distance-redshift relation, in order words, the Hubble law. The most often technique is parallax, i.e. use  geometry to calibrate the luminosity of pulsating Cepheid variables, for instance, which can be seen at great distances and thus allow measurements of the cosmic expansion. We highlight that this method treats such stars as standard candles. 

Hence, once those stars are empirically standardized, the same type has the same luminosity, without invoking any theoretical aspect. Though direct, this type of measurement is much more subject to systematic errors than the early universe measurements. Evidence of this fact is the last measurement presented in table (\ref{tab:my_label}). It uses Red Giants instead of Cepheids or Supernovae to evaluate the distance-redshift relation, and the value inferred for $H_0$ is a little bigger but compatible with the early universe evaluations, with the error bars. In many cases, the Red Giants and Cepheids used to obtain the different data sets are located in the same galaxies, the discrepancy in the results indicates the presence of a large systematic error in one or both of these measurements. New observations using the James Webb telescope may solve this issue in a couple of years, see a discussion of this conflict in Ref.~\cite{Freedman:2021ahq}.}

This cosmological problem has triggered several solutions, see for recent reviews \cite{DiValentino:2021izs,Kamionkowski:2022pkx}. One possibility to increase the Hubble constant inferred from CMB probes is to add some amount of radiation at early times. A plausible way to accomplish this is via the introduction of new light species that were in thermal equilibrium much before CMB decoupling \cite{Anchordoqui:2011nh,Boehm:2012gr,Brust:2013ova}. Such light species will contribute to the number of degrees of relativistic degrees of freedom, $N_{eff}$, which is positively correlated with the Hubble rate. An alternative way to increase $N_{eff}$ is to introduce a relativistic production mechanism of dark matter particles, which in turn mimic the effect of a neutrino species \cite{Hooper:2011aj}. 

It is well known that dark matter cannot be relativistic at matter-radiation equality for the sake of structure formation, and in this mechanism, dark matter is not being converted into dark matter radiation after Big Bang Nucleosynthesis \cite{Bringmann:2018jpr}. We are simply assuming that it was produced relativistically, but later it became non-relativistic much before Big Bang Nucleosynthesis or CMB decoupling. Therefore, this mechanism behaves just like any other standard non-relativistic dark matter model at late times. If a dark matter particle is produced relativisticaly it might be safely non-relativistic depending on when it was produced and its initial kinetic energy, as we will explain later. Anyway, this solution to the $H_0$ problem via $N_{eff}$ has proven to be insufficient with the latest data from Planck and new direct measurements of $H_0$. Within the $\Lambda$CDM one cannot find $H_0 > 70 km s^{-1} Mpc^{-1}$. Small deviations from the $\Lambda$CDM, however, allow larger values \cite{ref:Anchordoqui2021}. 

In the $\Lambda$CDM, the dark energy equation of state, $p = w \rho$, has $w=-1$, but in Phantom-like models $w < -1$ \cite{Caldwell:1999ew,Caldwell:2003vq,Vikman:2004dc,Nojiri:2005sx,Shafer:2013pxa,Ludwick:2017tox,Alestas:2020mvb}. This deviation in the equation of state allows larger values of $H_0$ when global fits to the CMB spectrum are performed \cite{ref:Anchordoqui2021}. It is known that Phantom-like cosmologies,  that experience late times dark energy transitions at redshifts $z\ll 0.1$ can raise the Hubble constant to values larger than $ 73{\rm  km s^{-1} Mpc^{-1}}$, while yielding equally good fit as $\Lambda$CDM at higher redshift data, in particular from the cosmic microwave background and baryon acoustic oscillations. Although, it faces some problems to raise $H_0$ to large values when data from SHOES collaboration \cite{Riess:2021jrx}, which consists in using Cepheid variables as intermediate calibrators, are accounted for \cite{Benevento:2020fev}. It has to do with the SNIa absolute magnitude obtained in these data analyses, which disagree with the absolute magnitude derived from
SNIa, BAO and CMB data \cite{Alestas:2020zol,Camarena:2021jlr}. Although, if a redshift dependence on the supernova absolute magnitude is included, the Phantom-like solution to the $H_0$ problem remains viable \cite{Alestas:2020zol,Camarena:2021jlr}. As Phantom-like cosmology still stands as a plausible solution to the $H_0$ problem, we will consider it as our cosmological model, allowing us to connect the increase in the Hubble rate with $N_{eff}$ in terms of dark radiation.
 
In this work, the dark radiation arises via the relativistic production of dark matter that occurs through a decaying process, where a heavy particle $(\chi^\prime)$ decays into a dark matter particle $(\chi)$ plus a photon $(\gamma)$. That decay adds an amount of hot dark matter which behaves as dark radiation for a while but later becomes non-relativistic as its energy decreases with the expansion. Hence, in this way, the production of dark matter contributes to $N_{eff}$ and thus increasing $H_0$. As a side remark, note that if the results of Refs.~\cite{Freedman:2019jwv,Freedman:2021ahq} discussed previously are correct, namely, that $H_0 \approx 70 km s^{-1} Mpc^{-1}$, this non-thermal production mechanism of dark matter particles would be sufficient to reconcile CMB and late time measurements of $H_0$ without appealing to any Phantom physics.

We will explain the mechanism in a model independent way, and later we write down non-renormalizable operators encompassing spin 0, spin 1 and spin 1/2 dark matter particles that feature this non-thermal production of dark matter particles. The important quantities are the masses of the particles and the energy scale of the effective operator, $\Lambda$. With this at hand, we delimit the region of parameter space iwhich offers a solution to the $H_0$ trouble.

This work is structured as follows: In section II we explain how the non-thermal production of dark matter raises $H_0$; In section III, we present effective operators that give rise to the $\chi^\prime \rightarrow \chi \gamma$ decay, and derive the corresponding decay width; In section IV we discuss the results, before concluding in section V.

\section{Increase in relativistic energy density produced by dark matter}

We are considering a radiation era where only photons and neutrinos are relativistic. Therefore, the total energy density in this stage is,
\begin{equation}
        \rho = \rho_{rad} = \frac{\pi^2}{30}g_*T^4,
\end{equation}
where $T$ is the temperature of the photons and $g_*$ is the total relativistic degrees of freedom \cite{piattella2018lecture}. The factor $g_*$ gives,
\begin{equation}
    g_* = g_\gamma + \frac{7}{4} \, g_{\nu}N_{\nu} \left( \frac{T_{\nu}}{T_{\gamma}} \right)^4 = 2 + \frac{}{} \frac{7}{4} \left( \frac{4}{11} \right)^{4/3} N_{\nu},
\end{equation}
where $g_\gamma = 2$ indicates that photons have two polarization states, $g_{\nu} = 1$ informs that standard model neutrinos are only left-handed, $T_\nu / T_{\gamma} = (4/11)^{1/3}$ is the ratio between the neutrinos and photons temperature after the neutrinos decoupling \cite{piattella2018lecture}, and $N_\nu$ is the number of neutrino flavors.

In the standard model, there are three neutrinos specie. Thus we expect $N_{eff}$ to be close to three, not precisely three, because of some temperature dependence. However, in non-standard cosmologies, we generally write $N_{eff} = 3 + \Delta N_{eff}$, where $\Delta N_{eff}$ refers to the extra number of relativistic degrees of freedom, which may come in sort of new light species or other mechanisms that mimic this effect. Denoting the energy density of a single standard model neutrino species as $\rho_{1\nu}$, we define,

\begin{equation}
    \Delta N_{eff} = \frac{\rho_{extra}}{\rho_{1\nu}}\cdot
    \label{eq:deltaN_general}
\end{equation}

Notice that the ratio between one neutrino species and cold dark matter energy density in the matter-radiation equality is,
\begin{equation}
    \left. \frac{\rho_{1\nu}}{\rho_{CDM}} \right|_{t = t_{eq}} = \frac{\Omega_{\nu,0}\rho_c}{3a^4_{eq}} \times \left( \frac{\Omega_{CDM,0}}{a^3_{eq}}  \right)^{-1} = 0.16,
    \label{eq:0.16}
\end{equation} where $\Omega_{\nu,0} = 3.65 \times 10^{-5}$, $\Omega_{CDM,0} = 0.265$, and $a_{eq} = 3 \times 10^{-4}$ \cite{Aghanim:2018eyx}.

Consequently, one neutrino density energy is equivalent to $16\%$ of the cold dark matter energy \cite{Alcaniz:2019kah} at the matter-radiation equality. In other words, if a fraction of dark matter is relativistic at that time, it can contribute to the energy density just like a neutrino species.  That said, we consider a heavy particle $\chi^\prime$ which decays in the radiation era in two particles, dark matter $(\chi)$ and a photon $(\gamma)$. We also assume that $m_{\chi^\prime} \gg m_{\chi}$, because we need dark matter to be produced relativistically so it can mimic the effect of a neutrino species. We avoid problems with structure formation by assuming that only a small fraction of dark matter particles are produced in this way \cite{Alcaniz:2019kah}. We will now devote some time explaining how this decaying process can generate $\Delta N_{eff}\neq 0$.

In the $\chi^\prime$ resting frame, the four-momentum of the particles are,
\begin{gather*}
    p_{\chi^\prime} = (m_{\chi^\prime}, \bm{0}),\\
    p_{\chi} = (E_{\chi}(\bm{p}), \bm{p}),\\
    p_{\gamma,\, \nu} = (|\bm{p}|, -\bm{p}).
\end{gather*}

Imposing four-momentum conservation we obtain, 
\begin{gather}
    \left|\bm{p}_{\chi}(\tau) \right| = \left|\bm{p} \right| = \frac{1}{2} m_{\chi^\prime} \left[ 1 - \left(\frac{m_{\chi}}{m_{\chi^\prime}} \right)^2 \right], \label{eq:momentum_chi}\\
    E_{\chi}(\tau) = m_{\chi} \left( \frac{m_{\chi^\prime}}{2m_{\chi}} + \frac{m_{\chi}}{2m_{\chi^\prime}} \right), \label{eq:energy_in_tau}
\end{gather} where $\tau$ is the $\chi^\prime$ lifetime. Note that the equation above refers to the energy and momentum at the moment immediately after the decay. Hence, the Lorentz factor is,
\begin{equation}
    \gamma_{\chi}(\tau) = \left( \frac{m_{\chi^\prime}}{2m_{\chi}} + \frac{m_{\chi}}{2m_{\chi^\prime}} \right),
\end{equation}where $E_{\chi}(\tau) = m_{\chi}\gamma_{\chi}(\tau)$.

The momentum of the particle is inversely proportional to the scale factor, thus $\bm{p}^2_{\chi} \propto \frac{1}{a^2}$, which implies in,
\begin{equation}
    \begin{split}
        &E^2_{\chi} - m^2_{\chi} = \bm{p}^2_{\chi} \propto \frac{1}{a^2}\\
        &\Rightarrow  \left( E^2_{\chi}(t) - m^2_{\chi} \right)a^2(t) = \left( E^2_{\chi}(\tau) - m^2_{\chi} \right)a^2(\tau)\\
        &\Rightarrow E_{\chi}(t) = m_{\chi}\left[ 1 + \left( \frac{a(\tau)}{a(t)} \right)^2 \left( \gamma^2_{\chi}(\tau) - 1\right)\right]^{1/2}.
        \label{eqEchi}
    \end{split}
\end{equation}

From Eq.\ref{eqEchi} we can extract the Lorentz factor for the dark matter particles at a given time $t$. Since we are considering a phase where the universe is radiation dominated, we can substitute $a(\tau)/a(t)$ for $\sqrt{\tau/t}$ and find the Lorentz factor \cite{ref:hobson2006GR}, 
\begin{equation}
    \gamma_{\chi}(t) = \sqrt{\frac{ (m^2_{\chi} - m^2_{\chi'})^2 }{4m^2_{\chi}m^2_{\chi'}} \left( \frac{\tau}{t} \right) + 1 }.
\end{equation}

A particle in the non-relativistic regime has its mass as the mean contribution of the total energy. Hence, the dark matter energy can be written as,
\begin{equation}
    E_{\chi} = m_{\chi}(\gamma_{\chi} -1) + m_{\chi}.
    \label{eqEchi}
\end{equation}
This equation provides a direct interpretation of $m_{\chi}(\gamma_{\chi} -1)$ as the mean contribution part of the particle energy in the ultrarelativistic regime. Therefore, the total energy of dark matter particles is given by the energy of the cold dark matter plus the energy of the relativistically produced dark matter component,
\begin{equation}
    E_{DM} = N_{HDM}m_{\chi}(\gamma_{\chi} -1) + N_{CDM}m_{\chi},
\end{equation}
where $N_{HDM}$ and $N_{CDM}$ are the number of hot and cold dark matter particles, respectively. To avoid conflicts with results from standard cosmology, we enforce $N_{HDM} \ll N_{CDM}$, as will explain further.

The ratio between the hot and cold dark matter energy density is
\begin{equation}
    \frac{\rho_{HDM}}{\rho_{CDM}} = \frac{n_{HDM}m_{\chi}\left( \gamma_{\chi} -1 \right)}{n_{CDM}m_{\chi}} \equiv f\left( \gamma_{\chi} -1 \right),
\end{equation}
where $n_{HDM}$ and $n_{CDM}$ are the number density of relativistic and nonrelativistic produced dark matter particles, respectively. The factor $f$ is the ratio between these two number densities and it must be small. Here, we consider $f = 0.01$, which is an upper limit obtained from structure formation \cite{Kelso:2013paa}. Note that this relativistically produced dark matter will be eventually cold, i.e. with $E_{\chi} \sim m_{\chi}$ shortly after structure formation begins.

We assume that the extra source of radiation in \eqref{eq:deltaN_general} is the dark matter particles in a hot stage. Using Eq.\ref{eq:deltaN_general}, we get,
\begin{equation}
    \Delta N_{eff} = \frac{\rho_{HDM}}{\rho_{1\nu}} = \frac{\rho_{CDM}f(\gamma_{\chi} - 1)}{\rho_{1\nu}}\cdot
\end{equation}

Our next step is to calculate this expression at matter-radiation equality, where $\rho_{CDM}/\rho_{1\nu}=1/0.16$,
\begin{equation}
    \Delta N_{eff} = \lim_{t \to t_{eq}} \frac{f\left( \gamma_{\chi} -1 \right)}{0.16}\cdot
    \label{eqNefffull}
\end{equation}

In the limit $m_{\chi^\prime} \gg m_{\chi}$, 
 we can simplify Eq.\eqref{eqNefffull} to,
\begin{equation}
    \gamma_{\chi}(t_{eq}) -1 \approx \gamma_{\chi}(t_{eq}) \approx \frac{m_{\chi^\prime}}{2m_{\chi}} \sqrt{\frac{\tau}{t_{eq}}}\cdot
\end{equation}which leads to,
\begin{equation}
        \Delta N_{eff} \approx 2.5 \times 10^{-3}\sqrt{\frac{\tau}{10^{6}s}} \times f\frac{m_{\chi'}}{m_{\chi}},
    \label{eq:deltaN}
\end{equation}where we used $t_{eq} \approx 50 000 ~ \text{years} \approx 1.6 \times 10^{12} ~s$ \cite{tongLecturesCosmology}.

The $\Delta N_{eff}$ is a function of four parameters: (i) the lifetime and (ii) the mass of $\chi^\prime$; (iii) the mass of $\chi$; (iv) the fraction of hot dark matter particles $(f)$, that we assume to be $0.01$. We will address this assumption in the next section.

\begin{figure*}[htb!]
    \centering
    \subfigure[]{
    \includegraphics[width = 0.95\columnwidth]{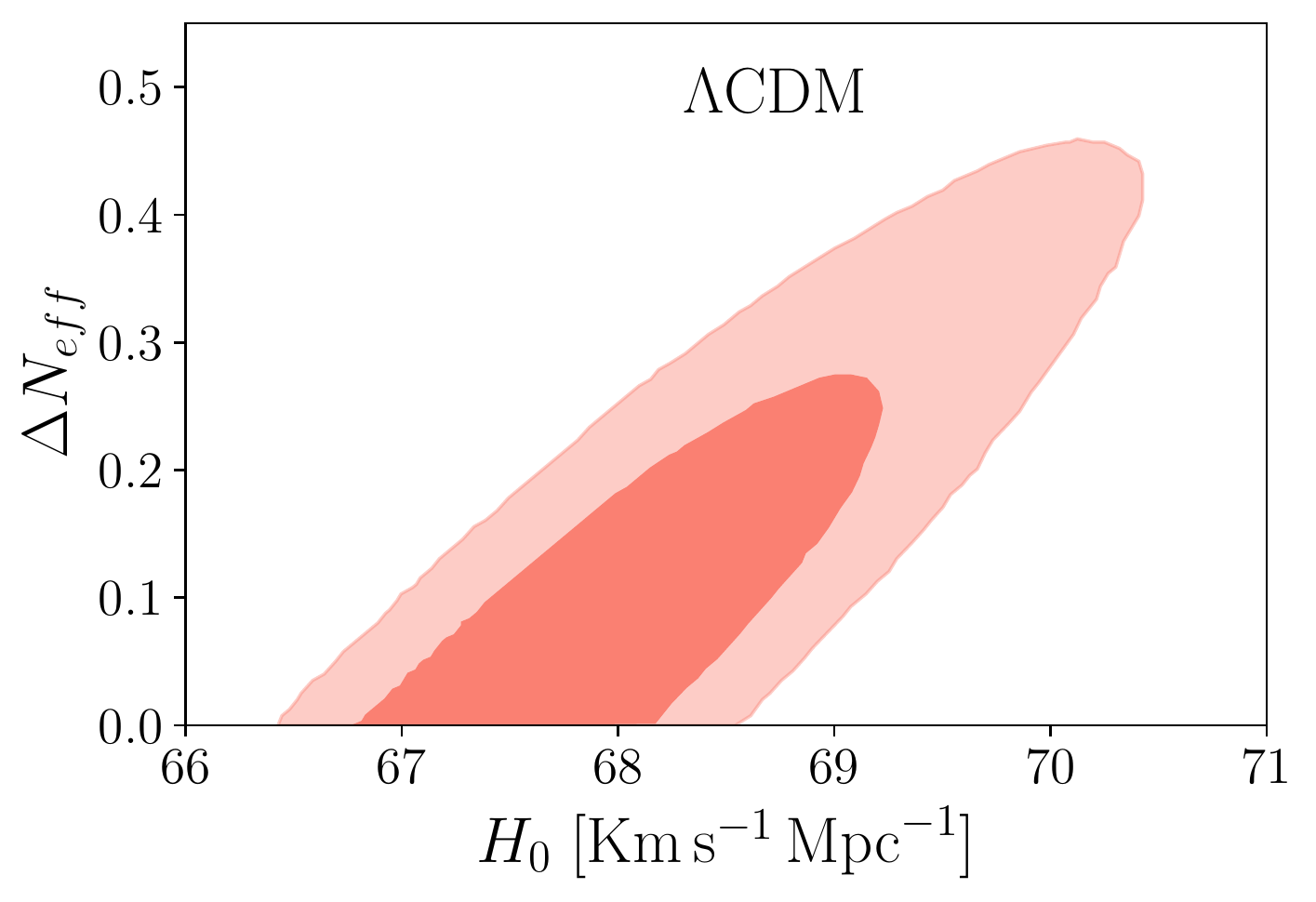}
    \label{fig:H0xNeff_LCDM}
    }
    \subfigure[]{
    \includegraphics[width = 0.95\columnwidth]{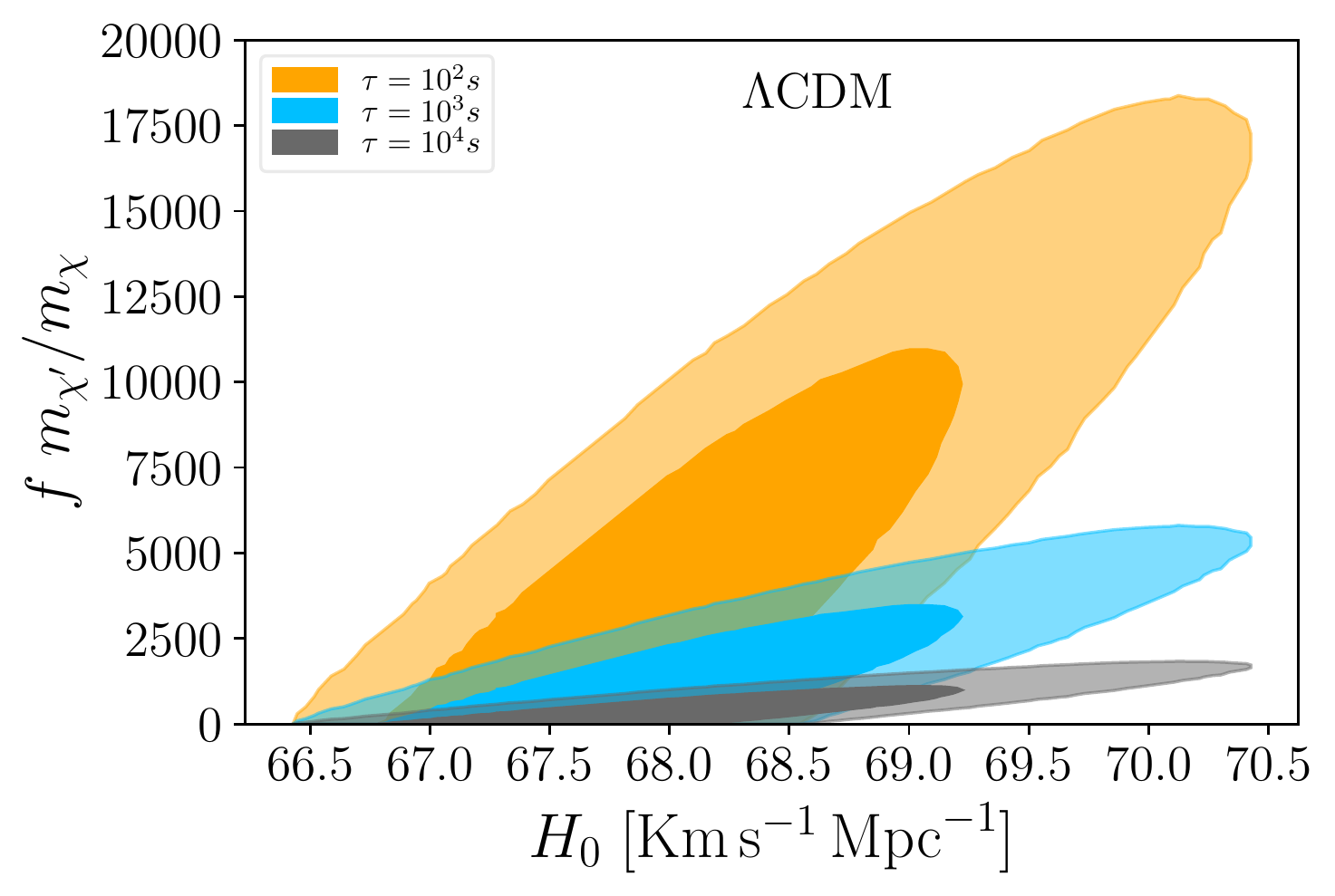}
    \label{fig:fxH0_LCDM}
    }
    \subfigure[]{
    \includegraphics[width = 0.95\columnwidth]{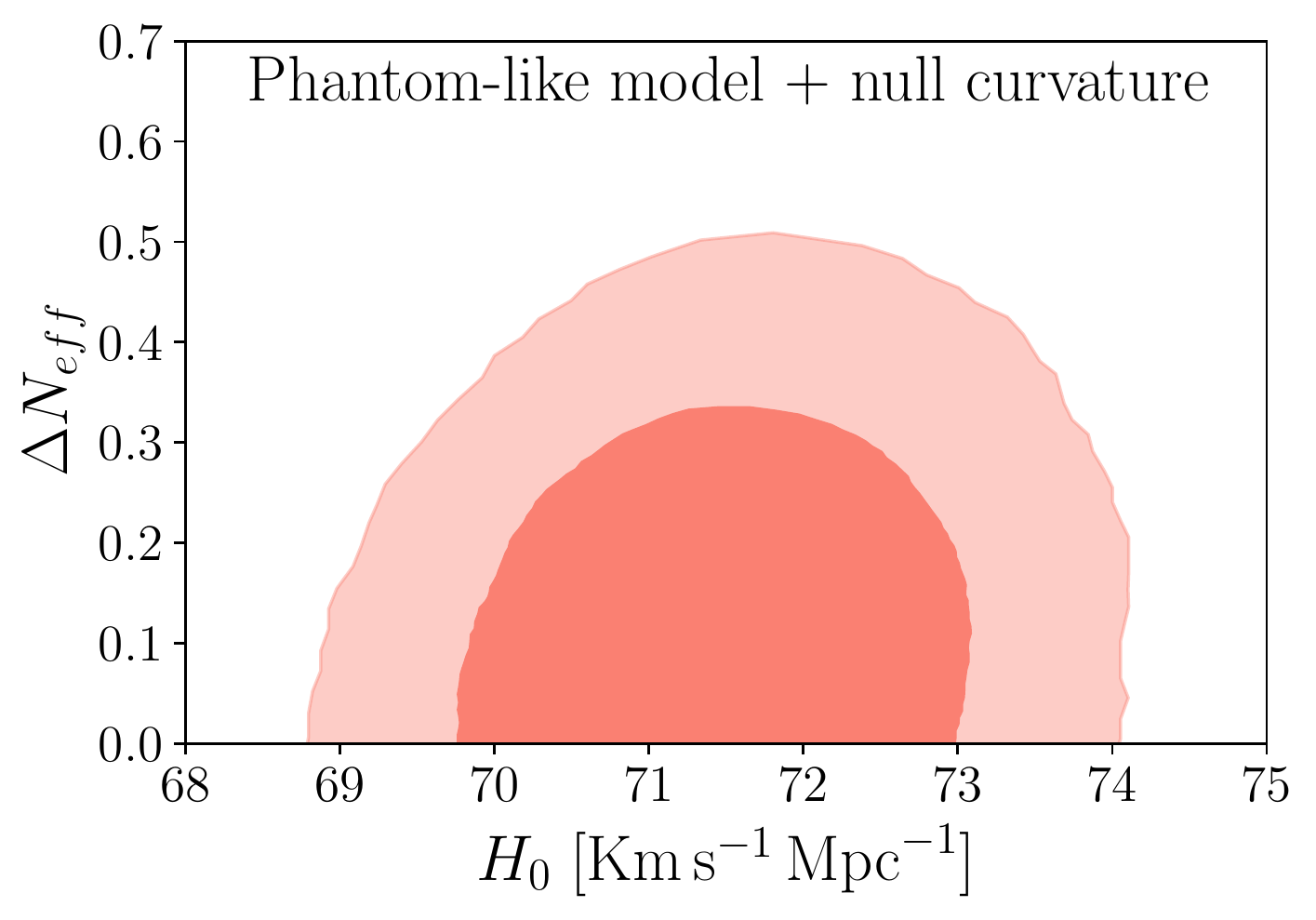}
    \label{fig:H0xNeff_nullCurvature}
    }
    \subfigure[]{
    \includegraphics[width = 0.95\columnwidth]{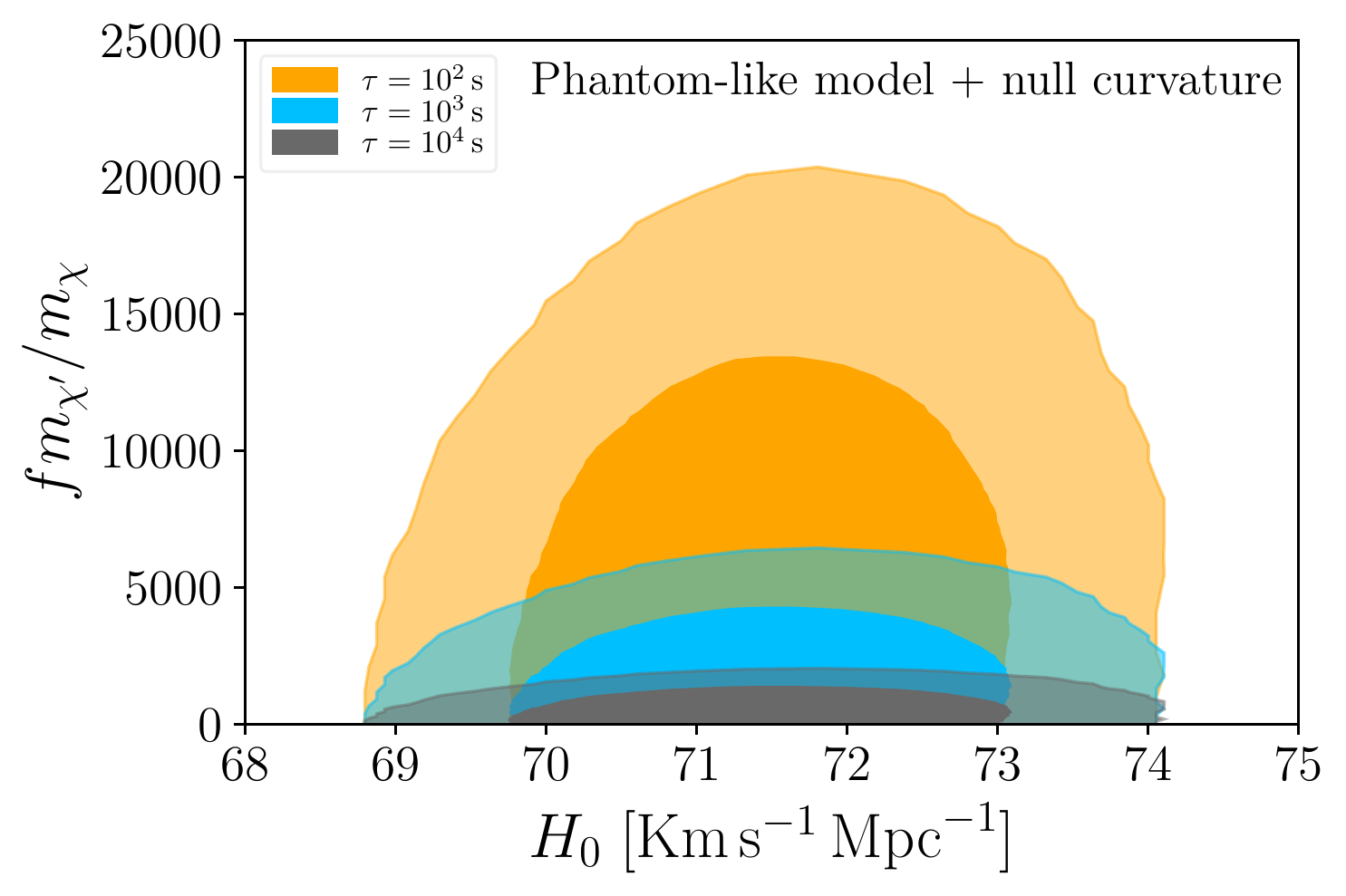}
    \label{fig:fxH0_nullCurvature}
    }
    \subfigure[]{
    \includegraphics[width = 0.95\columnwidth]{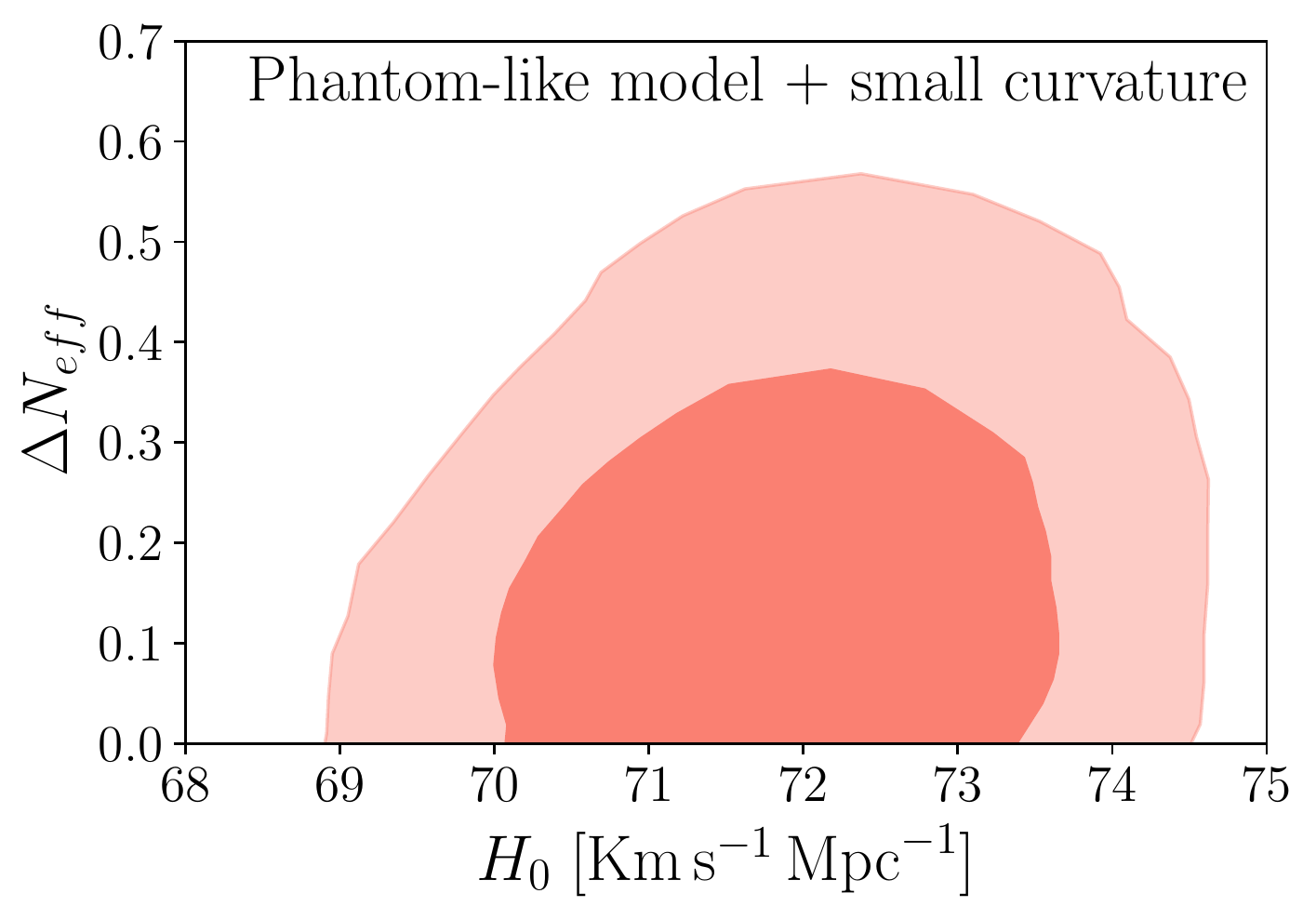}
    \label{fig:H0xNeff_smallCurvature}
    }
    \subfigure[]{
    \includegraphics[width = 0.95\columnwidth]{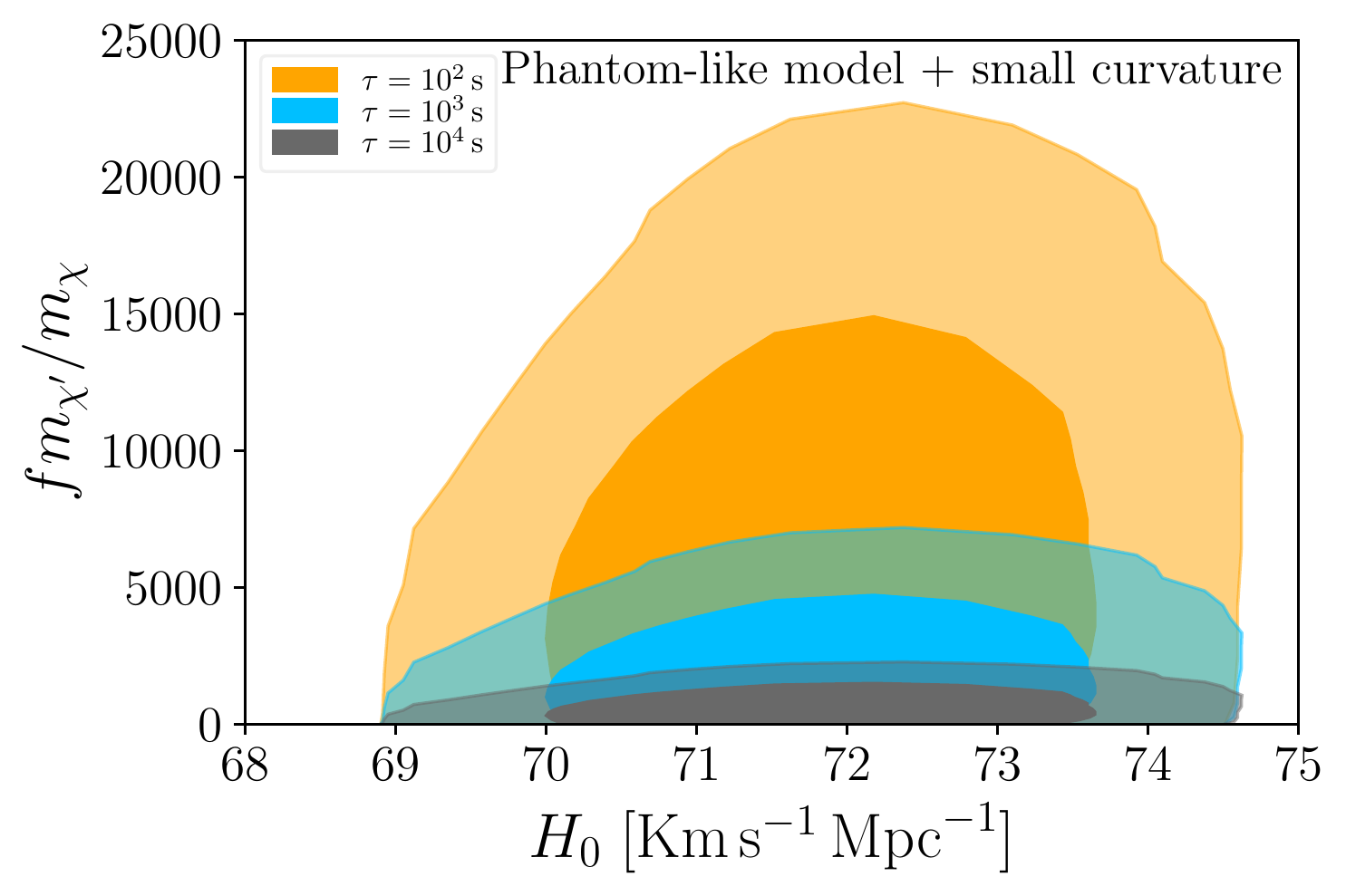}
    \label{fig:fxH0_smallCurvature}
    }
     \caption{Allowed regions of parameters that connect our mechanism and the value of Hubble constant in phantom-like cases. The first row corresponds to the $\Lambda CDM$ model, and in the second and third rows a phantom-like quintessence is introduced, first in a spatially flat model, then with non-null spatial curvature. The second column presents the cases corresponding to a non-zero $\Delta N_{eff}$. The data set that connects $\Delta N_{eff}$ and $H_0$ showed in \textbf{(a)}, \textbf{(c)}, and \textbf{(e)} is taken from \cite{ref:Anchordoqui2021}. In all figures, the lighter regions correspond to $99\%$ of CL, while the darkest regions correspond to $68\%$ of CL. In \textbf{(b)}, \textbf{(d)}, and \textbf{(f)} the orange, blue and gray regions correspond to the cases where $\chi^\prime$ lifetime is $10^2 s$, $10^3 s$ and $10^4 s$ respectively. The bounds use Planck 2018 CMB data, BAO, and type Ia data from the Pantheon sample.}
    \label{fig:fxH0}
\end{figure*}

As aforementioned, Planck collaboration has reported that $N_{eff}$ and $H_0$ are positively correlated. Therefore, we can use Eq.\eqref{eq:deltaN} to connect $H_0$ with $f m_{\chi^\prime}/m_{\chi}$  for a given lifetime.

Assuming that the Hubble constant measured locally should indeed be larger than $70\,  \text{Km} \, \text{s}^{-1} \, \text{Mpc}^{-1}$,  one can conclude that the 
$\Lambda CDM$ model does not suffice \cite{ref:Anchordoqui2021}. It is necessary to consider non-standard cosmological scenarios \cite{ref:Anchordoqui2021}. Here we will consider Phantom-like models \cite{Yang:2018euj}. We use the positive correlation between $H_0$ and $N_{eff}$ found in \cite{ref:Anchordoqui2021}, and derive the allowed values of $H_0$  for choices of the product $fm_{\chi^\prime} / m_{\chi}$ for $\tau= 10^2$~s, $10^3$~s, and $10^4$~s. 

Phantom-like cosmologies alone allow $H_0$ values larger than  $70\,  \text{Km} \, \text{s}^{-1} \, \text{Mpc}^{-1}$, and consequently can solve the $H_0$ trouble. This explains why in Fig.\ref{fig:fxH0} $fm_{\chi^\prime} / m_{\chi}$ can go to zero. However, if we adopted local measurements pointing to $H_0\leq 70\,  \text{Km} \, \text{s}^{-1} \, \text{Mpc}^{-1}$, there would be no need for a Phantom-like cosmology because our mechanism of non-thermal production of dark matter particles can yield $N_{eff}=3.3$ and, consequently,  $H_0= 70\,  \text{Km} \, \text{s}^{-1} \, \text{Mpc}^{-1}$ \cite{ref:Anchordoqui2021}.  In  the middle panel of Fig.\ref{fig:fxH0} we assumed null curvature and in the bottom a non-zero curvature. The difference in the parameter space is mild. Thus, we can safely say that with or without curvature our work can solve the $H_0$ trouble. 

We stress that the advantage of our mechanism is the interplay between particle physics and cosmology. Instead of relying simply on a cosmological model such as Phantom-like cosmology, our idea invokes a connection to the dark matter density and to the production mechanism of dark matter particles. As far as typical direct detection searches go \cite{XENONCollaboration:2022kmb}, dark matter particles with a non-thermal origin in the early universe produce no effect on the scattering rate observed today. However, if dark matter particles experienced in the early universe a non-thermal production, the parameter space probed by direct detection experiments, in terms of mass and coupling of given model, changes. Hence, If only a small fraction of dark matter is produced non-thermally, this brings no impact to the typical direct detection or accelerator searches for dark matter particles \cite{Duarte:2022feb}. Be that as it may, this small fraction might serve an interesting purpose in cosmology, a solution to the $H_0$ problem. 
We will now address some important cosmological aspects of this non-thermal production mechanism in the early universe. We start discussing structure formation and Big Bang Nucleosynthesis.

\section{Relevant Bounds}

\subsection{Big Bang Nucleosynthesis}

When electromagnetic energy is injected in to the universe through non-thermal processes as the one we are considering Double Compton scattering ($\gamma e^-\rightarrow \gamma\gamma e^-$), and bremsstrahlung ($e^-X \rightarrow e^-X\gamma$) may alter the CMB spectrum \cite{Feng:2003uy,Feng:2004zu}, relaxing it to a Bose-Einstein distribution function with chemical potential different from zero. Given the existing upper limit on the chemical potential, we can limit the energy injection at a given time. The bounds are rather stringent, but for $\tau> 10^4$~s. In our work, we will focus on the region of parameter space in which $\tau< 10^4$~s, to avoid conflicts with BBN \cite{Hooper:2011aj,Alcaniz:2022oow}.

\subsection{Structure Formation}

Galaxy cluster observations restrict the amount of hot dark matter in the universe. Hot dark matter is typically treated as massive neutrinos. Those studies limit the fraction of hot dark matter in the universe, $\Omega_{HDM} /\Omega_{CDM}$ to be less than $0.01$. For this reason, we will consider $f=0.01$. Notice that we are being very conservative by taking this bound at face value, because the dark matter particles can be heavy, conversely to neutrinos. Therefore, its free-streaming evolves differently. A more robust calculation would have to be derive a more precise constraint.  

\subsection{Energy Evolution of Dark Matter}

In our formalism, a fraction of dark matter particles are created in a hot stage. But it is important that at matter-radiation equality time their kinetic energy had been lost, due to the expansion of the universe. The evolution of the dark matter particles are computed using Eq.\eqref{eqEchi}. Therefore, we can assess whether the dark matter particles produced this way are non-relativistic, i.e, $E_\chi \sim m_{\chi}$, at the matter-radiation equality. Focusing only on the region of parameter space which solves the $H_0$ we compute the dark matter energy at the matter-radiation equality. 

\begin{figure}[htb!]
    \centering
    \subfigure[]{
    \includegraphics[width=\columnwidth]{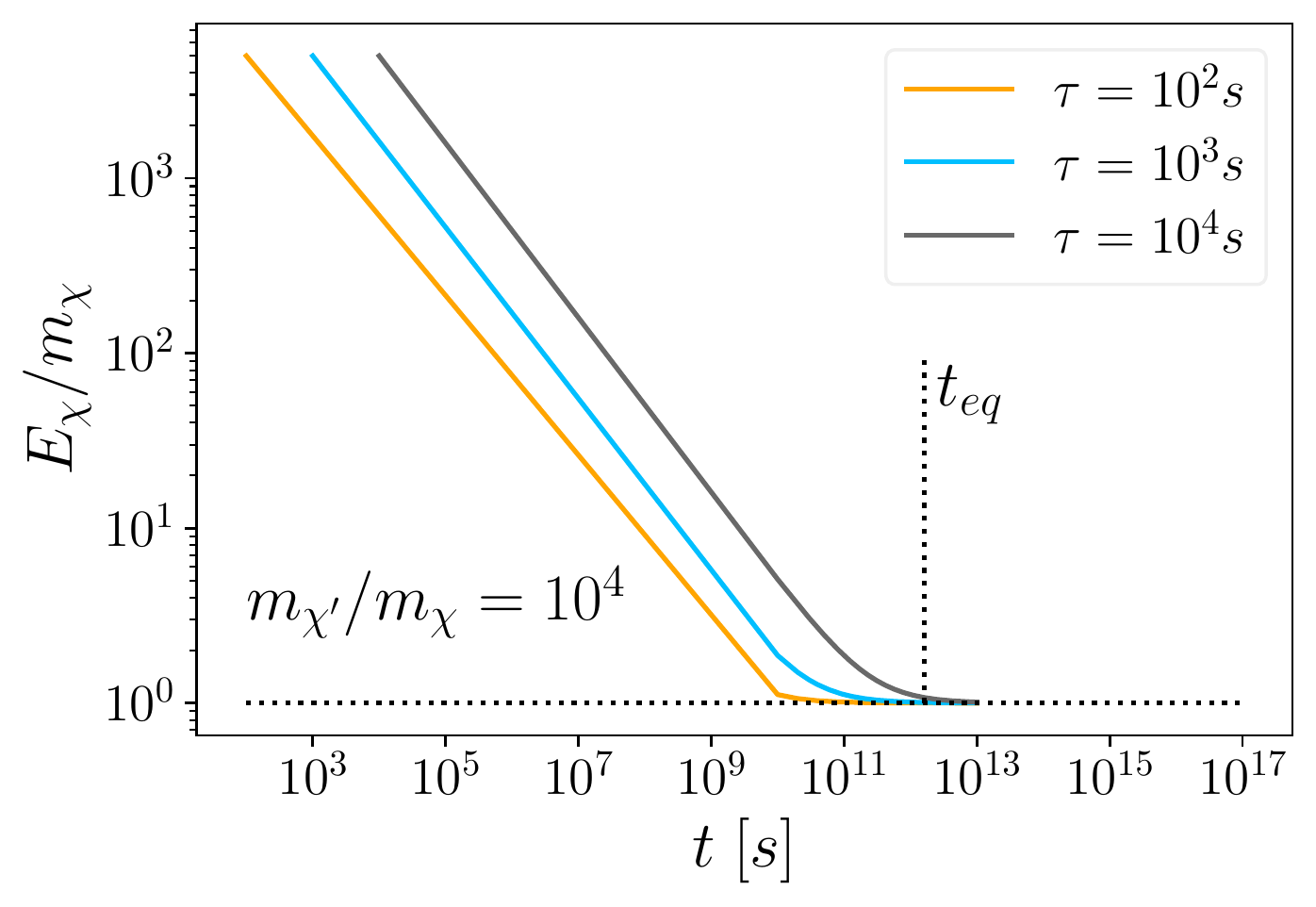}
    \label{fig:time_x_energy_caseE4}
    }
    \subfigure[]{
    \includegraphics[width=\columnwidth]{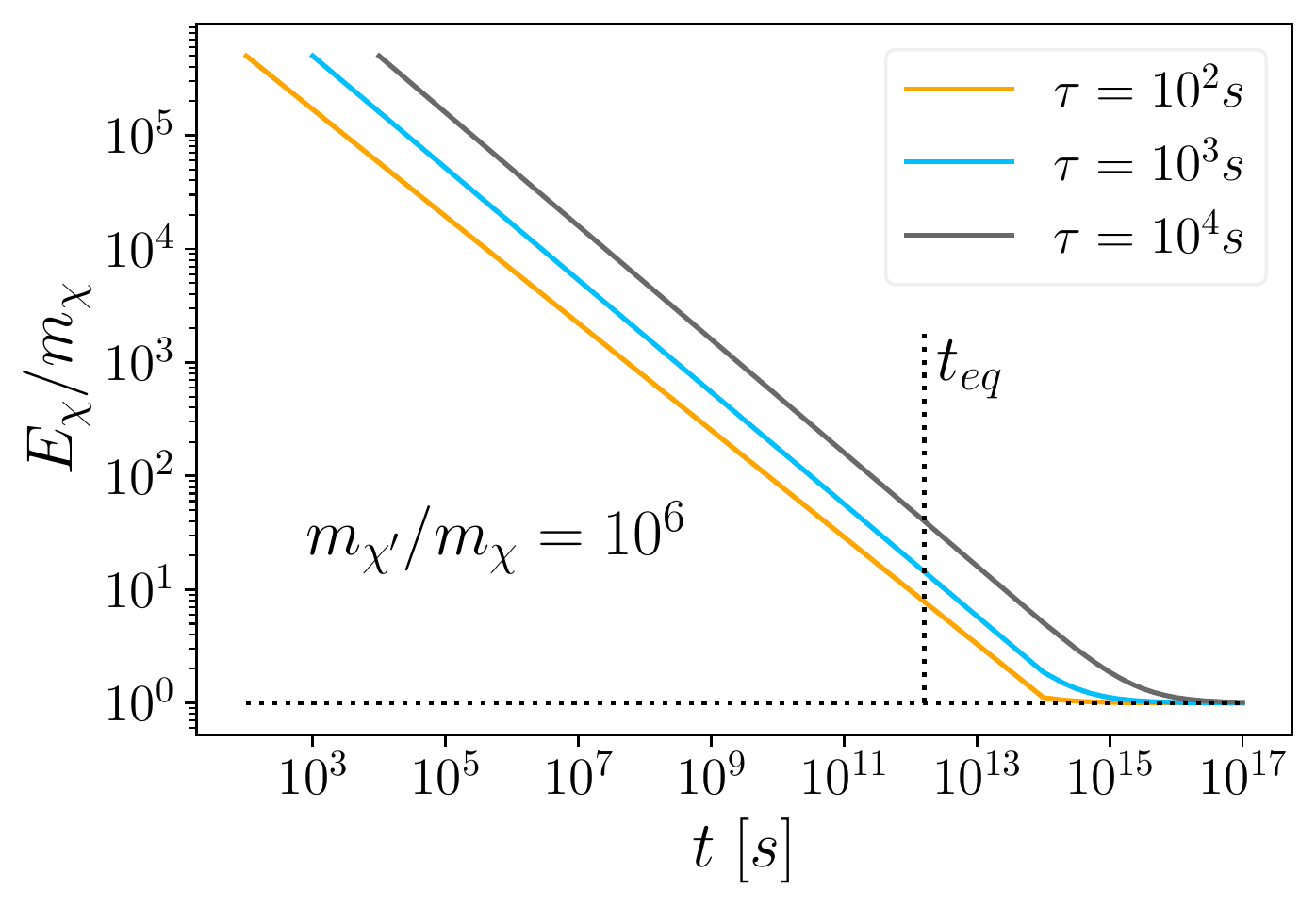}
    \label{fig:time_x_energy_caseE6}
    }
    \caption{Time evolution of dark matter energy. We consider situations where dark matter mother lifetime is $\tau = \{10^2 s, 10^3 s, 10^4 s\}$ and the ratio between dark matter mother and dark matter mass is $m_{\chi^\prime}/m_{\chi} = \{ 10^4, 10^6\}$. In all situations of \textbf{(a)}, dark matter is cold at matter-radiation equality $(t_{eq})$, while in \textbf{(b)} dark matter is hot in all scenarios.}
    \label{fig:time_x_energy}
\end{figure}

Taking $\tau \sim 10^2 s - 10^4 s$ and $m_{\chi^\prime}/m_{\chi} \sim 10^4 - 10^6$, which is within the region of interest to solve the $H_0$ problem, we show in Fig. \ref{fig:time_x_energy_caseE4} that the dark matter particles become non-relativistic at matter-radiation equality for $m_{\chi^\prime}/m_\chi = 10^4$. In Fig. \ref{fig:time_x_energy_caseE6} it is shown that for $m_{\chi^\prime}/m_\chi = 10^6$ dark matter particles are still relativistic at matter-radiation equality. Enforcing the dark matter particles to be cold at $t_{eq}$ we find a upper limit  on the mass ration $m_{\chi^\prime}/m_\chi$. We emphasize that this result is independent of $f$. We highlight that the choices for the parameter in the figures solve the $H_0$ discrepancy.  Despite the energy of dark mater being independent of $f$, we needed to assume $f$ to be small to reproduce the correct value of $H_0$. Therefore, changing $f$ means changing the lifetime and mass ratio that yields the correct $H_0$. That would consequently change the curves in Fig. \ref{fig:time_x_energy_caseE4} at matter-radiation equality for $m_{\chi^\prime}/m_\chi = 10^4$. Anyway, in Fig. \ref{fig:time_x_energy_caseE6}, our mechanism goes in the direction of a mixed cold+hot dark matter scenario, which may solve some small scale problems appearing in purely cold dark matter simulations \cite{Klypin:1992sf,Bose:2018juc,Lopez-Corredoira:2022dxt,Lahav:2022poa,ParticleDataGroup:2022pth}. 

In summary, our mechanism does not alter the CMB, BBN or structure formation prediction for the region of interest. We now move to a more particle physics-oriented section. Having in mind that this decay $\chi^\prime \rightarrow \chi +\gamma$ can solve the $H_0$ problem, we write down effective operators that feature this decay to determine the energy scale $\Lambda$ at which the $H_0$ can be solved through non-thermal production of dark matter particles.

\section{Effective theory of Dark Matter}

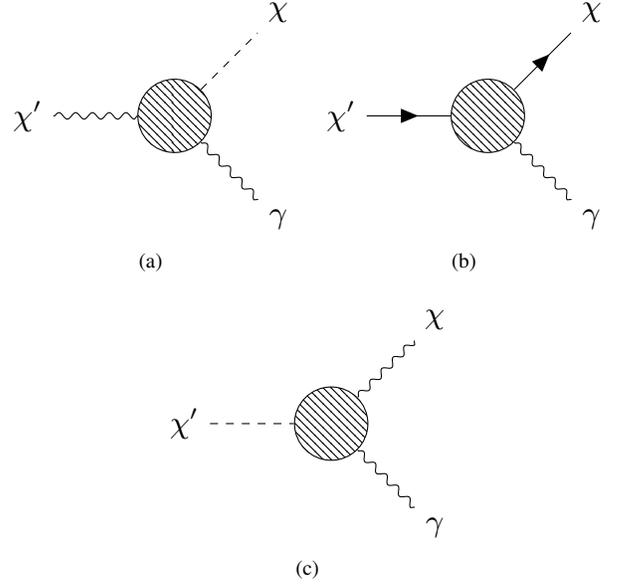
\begin{figure}[htb!]
    \centering
    \subfigure[]{
        \begin{tikzpicture}
            \begin{feynman}[scale = 1.3, transform shape]
                \vertex[blob] (a) {};
                \vertex [left = of a] (b) {$\chi^\prime$};
                \vertex [above right = of a] (c) {$\chi$};
                \vertex [below right = of a] (d) {$\gamma$};
                \diagram{
                    (b) --[boson] (a) --[scalar] (c);
                    (a) --[boson] (d);
                };
            \end{feynman}
        \end{tikzpicture}
        \label{fig:caseI_VSG}
    }
    \subfigure[]{
        \begin{tikzpicture}
        \begin{feynman}[scale = 1.3, transform shape]
            \vertex[blob] (a) {};
            \vertex [left = of a] (b) {$\chi^\prime$};
            \vertex [above right = of a] (c) {$\chi$};
            \vertex [below right = of a] (d) {$\gamma$};
            \diagram{
                (b) --[fermion] (a) --[fermion] (c);
                (a) --[boson] (d);
            };
        \end{feynman};
    \end{tikzpicture}
    \label{fig:caseII_FFG}
    }
    \subfigure[]{
        \begin{tikzpicture}
        \begin{feynman}[scale = 1.3, transform shape]
            \vertex[blob] (a) {};
            \vertex [left = of a] (b) {$\chi^\prime$};
            \vertex [above right = of a] (c) {$\chi$};
            \vertex [below right = of a] (d) {$\gamma$};
            \diagram{
                (b) --[scalar] (a) --[boson] (c);
                (a) --[boson] (d);
            };
        \end{feynman};
    \end{tikzpicture}
    \label{fig:caseIII_SVG}
    }
    \caption{Diagrammatic representation of a heavy particle $(\chi^\prime)$ that decay in hot dark matter $(\chi)$ and photon $(\gamma)$. Three cases are considered: \textbf{(a)} $\chi^\prime$ is spin-1 and $\chi$ is a spin-0 particle; \textbf{(b)} $\chi^\prime$ and $\chi$ are spin-1/2 particles; \textbf{(c)} $\chi^\prime$ is spin-0 and $\chi$ is a spin-1 particle.}
    \label{fig:diagrams}
\end{figure}

As the nature of dark matter is unknown, we will consider three effective operators of dimension five covering spin-0, spin-1 and spin-1/2 dark matter particles for the decay process $\chi^\prime \rightarrow \chi + \gamma$. The corresponding Feynmann diagrams are displayed in \fig{fig:diagrams}.

A two-body decaying rate is given by \cite{griffiths2020introduction},
\begin{equation}
    \Gamma(\chi^\prime \rightarrow \chi + \gamma) = \frac{\left|\bm{p}_{\chi} (\tau) \right|}{8\pi m^2_{\chi^\prime}} \left| \mathcal{M} \right|^2,
\end{equation} where $\mathcal{M}$ is the invariant amplitude. After plugging in the kinematics given in \eq{eq:momentum_chi}, we get,
\begin{equation}
    \Gamma = \frac{1}{16\pi m_{\chi^\prime}}\left[ 1 - \left(\frac{m_{\chi}}{m_{\chi^\prime}} \right)^2 \right] \left| \mathcal{M} \right|^2.
    \label{eq:gamma_general}
\end{equation}

We will use this general expression to calculate the lifetime $\tau = 1/\Gamma$ for three different effective operators presented below.

\subsection{Decay in spin-0 dark matter and photon}
In the first case we assume that $\chi^\prime$ is a spin-1, $\chi$ is a spin-0 particle, and the effective Lagrangian describing this decay $\chi^\prime \rightarrow \chi + \gamma$ is,
\begin{equation}
    \mathcal{L}_{eff} = \frac{1}{\Lambda} \phi_{\chi} \chi^\prime_{\mu \nu} F^{\mu \nu},
\end{equation}
where $\Lambda$ is an energy scale to be determined later. Note that $\chi^\prime_{\mu \nu} \equiv \partial_\mu \chi^\prime_\nu - \partial_\nu \chi^\prime_\mu$. The Feynman diagram for this process is shown in \fig{fig:caseI_VSG}, which results in,
\begin{equation}
    \left| \mathcal{M}\right|^2 = \frac{2m^4_{\chi^\prime}}{3\Lambda^2} \left[ 1 - \left( \frac{m_{\chi}}{m_{\chi^\prime}} \right)^2 \right].
\end{equation}
Substituting this result in \eq{eq:gamma_general} we obtain,
\begin{equation}
    \Gamma = \frac{m^3_{\chi^\prime}}{24\pi \Lambda^2} \left[ 1 - \left( \frac{m_{\chi}}{m_{\chi^\prime}}  \right)^2 \right]^3 \approx \frac{m^3_{\chi^\prime}}{24\pi \Lambda^2}\cdot
    \label{eq:gamma_VSG}
\end{equation}

\begin{figure}[htb!]
    \centering
    \subfigure[]{
    \includegraphics[width=\columnwidth]{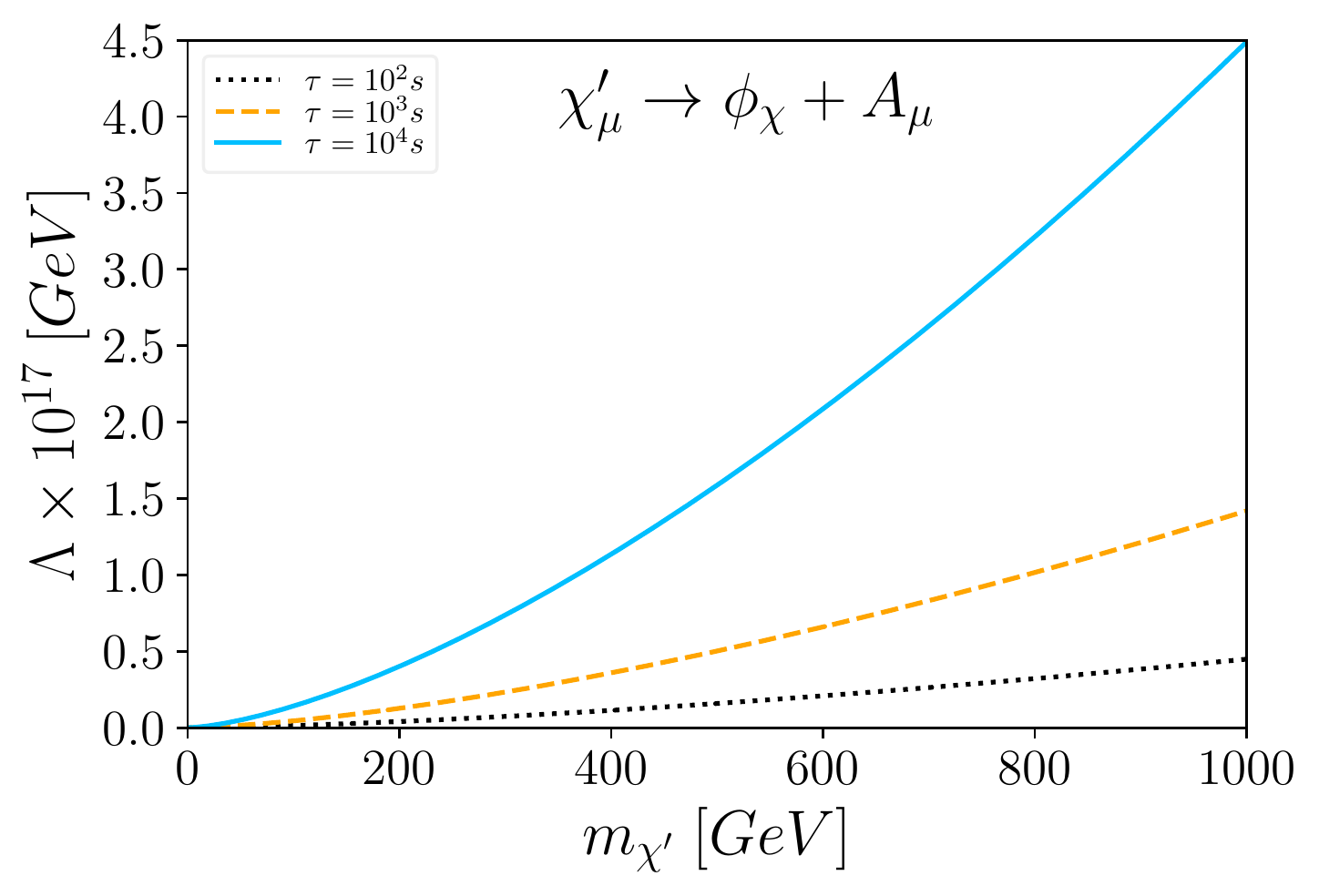}    
    \label{fig:MxLambda_case1_VSG}}        
    \subfigure[]{
    \includegraphics[width=\columnwidth]{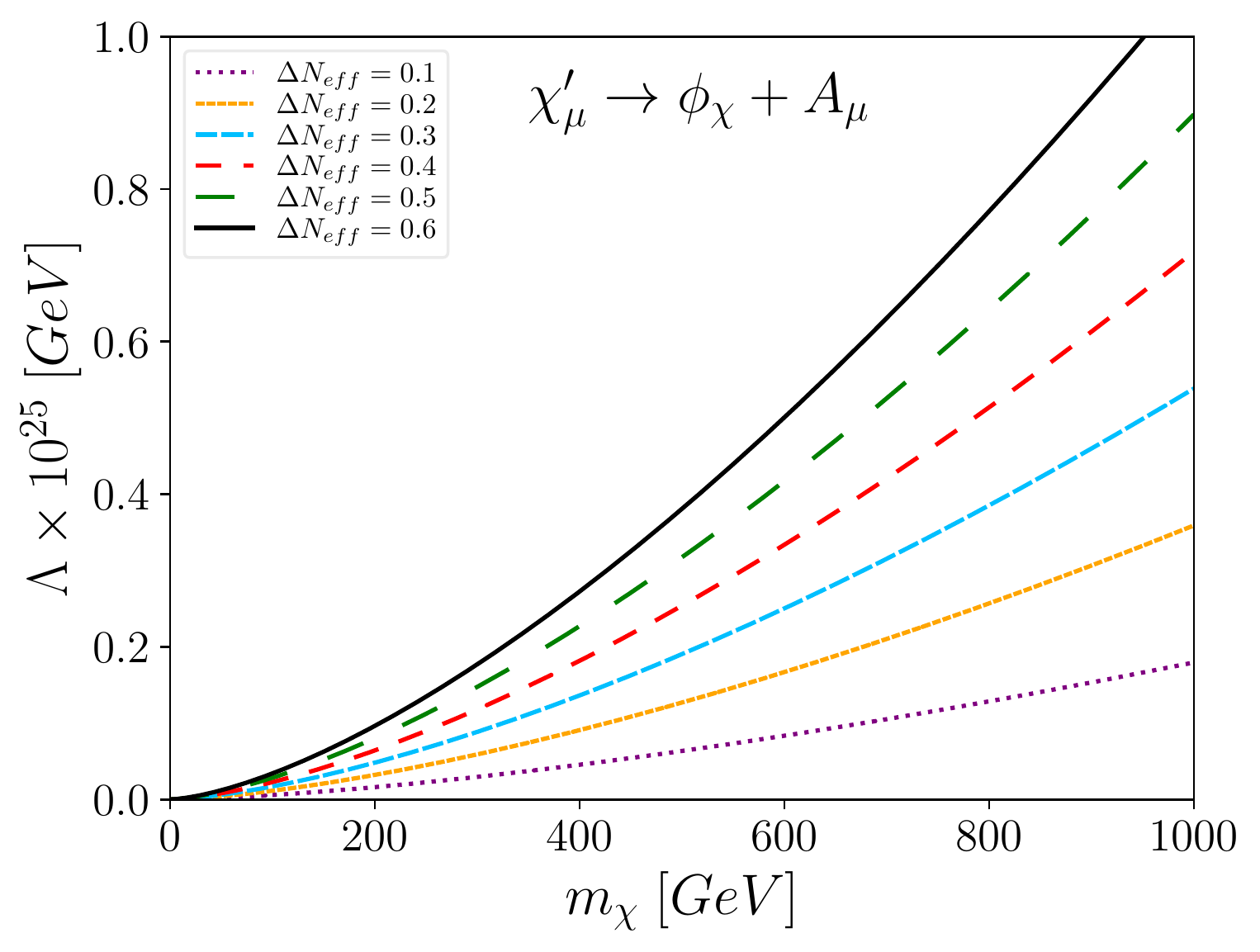}
    \label{fig:plot_Lambda_mchi_mod1}}
    \caption{Plot of $\Lambda$ as a function of the $\chi$ and $\chi^\prime$ masses for the case where $\chi^\prime$ is a spin-1, $\chi$ is a spin-0 particle. \textbf{(a)} $\Lambda \times m_{\chi^\prime}$ curves  built from Eq. (\ref{eq:gamma_VSG}), using $\tau = 10^2, 10^3 \, \text{and} \, 10^4 \,\text{s}$. \textbf{(b)} $\Lambda \times m_{\chi}$ curves constructed from Eqs. (\ref{eq:deltaN}) and (\ref{eq:gamma_VSG}). We consider the cases where $\Delta N_{eff} = 0.1 - 0.6$, with $f=0.01$ and $m_{\chi^\prime}/m_{\chi} = 10^4$.}
    \label{fig:plots_caseI}
\end{figure}

Therefore, the lifetime is set by $\Lambda$ and $m_{\chi^\prime}$. We exhibit this relation in \fig{fig:MxLambda_case1_VSG} for $\tau=10^2~s,10^3~s,10^4$~s. As $\Delta N_{eff}$ is now a function of $m_{\chi^\prime}$, $m_{\chi}$ and $\Lambda$, we can play with those quantities to outline the region of parameter space that solves the $H_0$ trouble exploiting its correlation with $\Delta N_{eff}$. In \fig{fig:plot_Lambda_mchi_mod1} we set $m_{\chi}/m_{\chi}=10^4$ and show the values of $\Lambda$ which yield $\Delta N_{eff}=0.1-0.6$ and lead to $H_0 \sim 70-72 {\rm km s^{-1} Mpc^{-1}}$ according to \fig{fig:fxH0_LCDM}. We would like to stress once more that if local measurement converge to  $H_0 \sim 70 {\rm km s^{-1} Mpc^{-1}}$, our mechanism alone is sufficient to solve the discrepancy on $H_0$, as can be seen in \fig{fig:H0xNeff_LCDM}. 

\subsection{Decay in spin-1/2 dark matter and  photon}

In the second possibility, we consider that $\chi^\prime$ and $\chi$ are spin-1/2 fermions, and the effective theory to describe the decay $\chi^\prime \rightarrow \chi + \gamma$ is
\begin{equation}
   \mathcal{L}_{eff} = \frac{1}{\Lambda}\bar{\psi}_{\chi}\sigma^{\mu \nu}\psi_{\chi^\prime} F_{\mu \nu} + h.c.,
\end{equation}
where $\sigma^{\mu \nu} = \frac{i}{2}[\gamma^\mu, \gamma^\nu]$. The corresponding Feynman amplitude is,
\begin{equation}
    \left| \mathcal{M}\right|^2 = \frac{8m_{\chi^\prime}^4}{\Lambda^2}\left[1- \left(\frac{m_\chi}{m_{\chi^\prime}}\right)^2\right]^2.
\end{equation}
Using \eq{eq:gamma_general} we find,
\begin{equation}
    \Gamma = \frac{m^3_{\chi^\prime}}{2\pi \Lambda^2} \left[ 1 - \left( \frac{m_{\chi}}{m_{\chi^\prime}}  \right)^2 \right]^3 \approx \frac{m^3_{\chi^\prime}}{2\pi \Lambda^2} \cdot
    \label{eq:gamma2_FFG}
\end{equation}

\begin{figure}[htb!]
    \centering
    \subfigure[]{
    \includegraphics[width=\columnwidth]{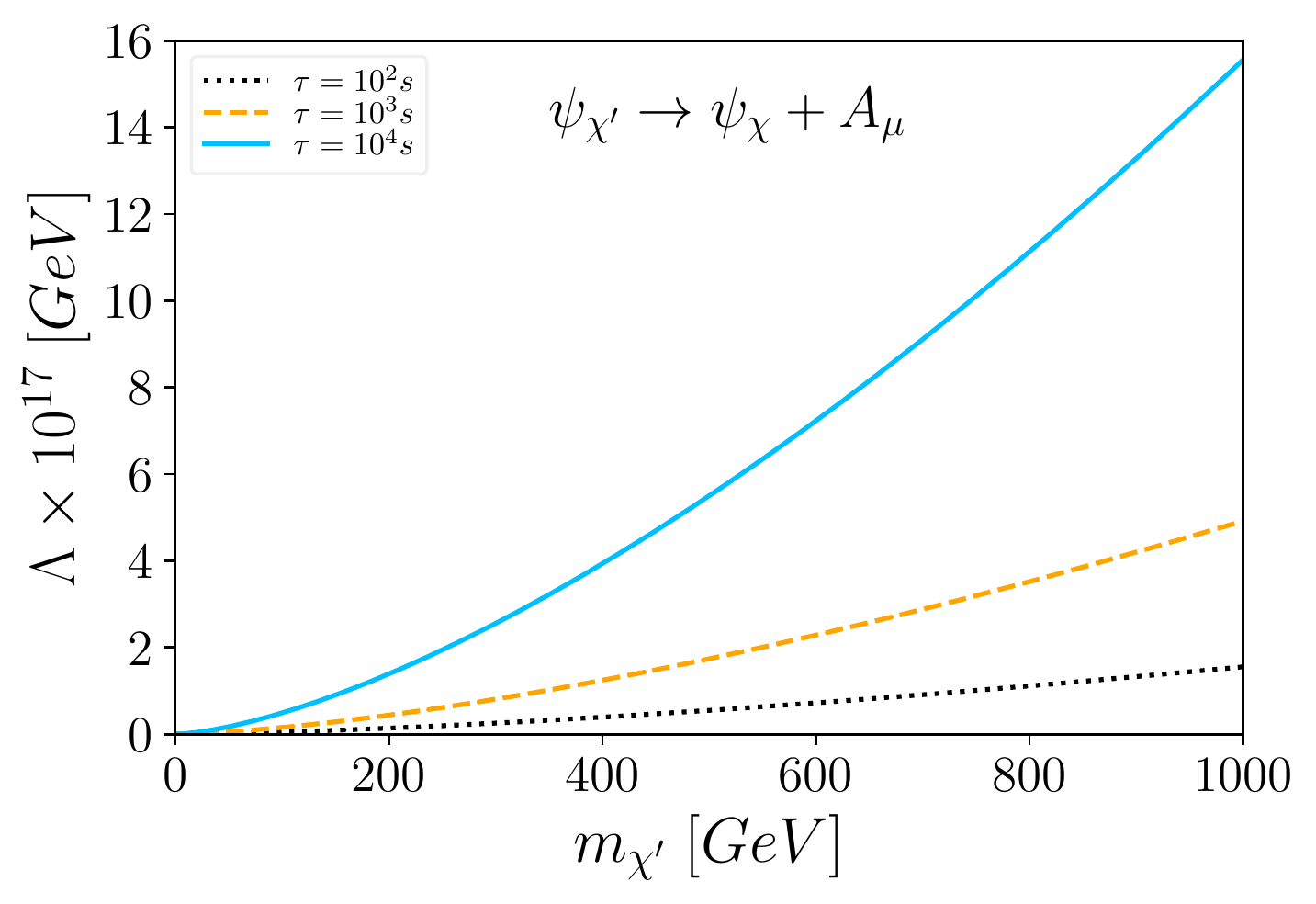}
    \label{fig:MxLambda_case2_FFG}}
    \subfigure[]{
    \includegraphics[width=\columnwidth]{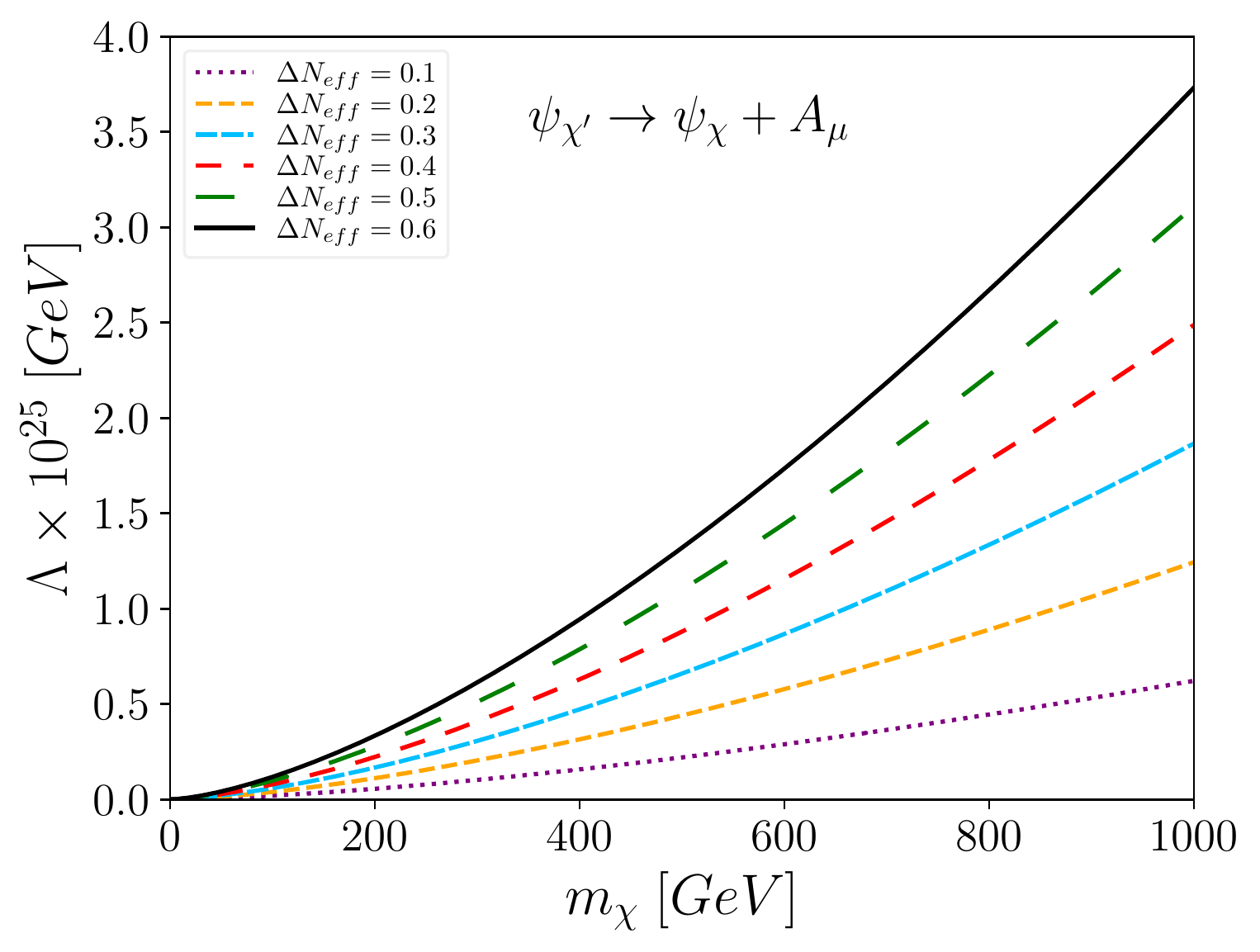}
    \label{fig:plot_Lambda_mchi_mod2}}
    \caption{Plot of $\Lambda$ as a function of the $\chi$ and $\chi^\prime$ masses for the case where $\chi^\prime$ and $\chi$ are spin-1/2 particles. \textbf{(a)} $\Lambda \times m_{\chi^\prime}$ curves  built from Eq. (\ref{eq:gamma2_FFG}), using $\tau = 10^2, 10^3 \, \text{and} \, 10^4 \,\text{s}$. \textbf{(b)} $\Lambda \times m_{\chi}$ curves constructed from Eqs. (\ref{eq:deltaN}) and (\ref{eq:gamma2_FFG}). We consider the cases where $\Delta N_{eff} = 0.1 - 0.6$, with $f=0.01$ and $m_{\chi^\prime}/m_{\chi} = 10^4$.}
    \label{fig:plots_caseII}
\end{figure}

In a similar vein, we use Eq.\eqref{eq:gamma2_FFG} to plot the relation between $\Lambda$ and $m_{\chi}$ for $\tau=10^2~s,10^3~s, 10^4~s$ in \fig{fig:MxLambda_case2_FFG}. Moreover, we derive the energy scale $\Lambda$ that reproduces $\Delta_{eff}=0.1-0.6$ and can lead to a solution to the $H_0$ trouble in \fig{fig:plot_Lambda_mchi_mod2}, assuming $m_{\chi^\prime}/m_{\chi}=10^4$, and $f=0.01$.

\subsection{Decay in spin-1 dark matter and photon}

Lastly, we take $\chi^\prime$ to be a spin-0 particle, $\chi$ is a spin-1/2 fermion, which is described by the effective operator, 

\begin{equation}
   \mathcal{L}_{eff} = \frac{1}{\Lambda}\phi_{\chi^\prime}\chi_{\mu \nu}F^{\mu \nu},
\end{equation}
where $\chi_{\mu \nu} \equiv \partial_\mu \chi_\nu - \partial_\nu \chi_\mu$, which yields,
\begin{equation}
    \left| \mathcal{M}\right|^2 = \frac{2m^4_{\chi^\prime}}{\Lambda^2}\left[ 1 - \left( \frac{m_{\chi}}{m_{\chi^\prime}} \right)^2 \right]^2,
\end{equation}and,
\begin{equation}
    \Gamma = \frac{m^3_{\chi^\prime}}{8\pi \Lambda^2} \left[ 1 - \left( \frac{m_{\chi}}{m_{\chi^\prime}}  \right)^2 \right]^3 \approx \frac{m^3_{\chi^\prime}}{8\pi \Lambda^2} \cdot
    \label{eq:gamma3_SVG}
\end{equation}

\begin{figure}[htb!]
    \centering
    \subfigure[]{
    \includegraphics[width=\columnwidth]{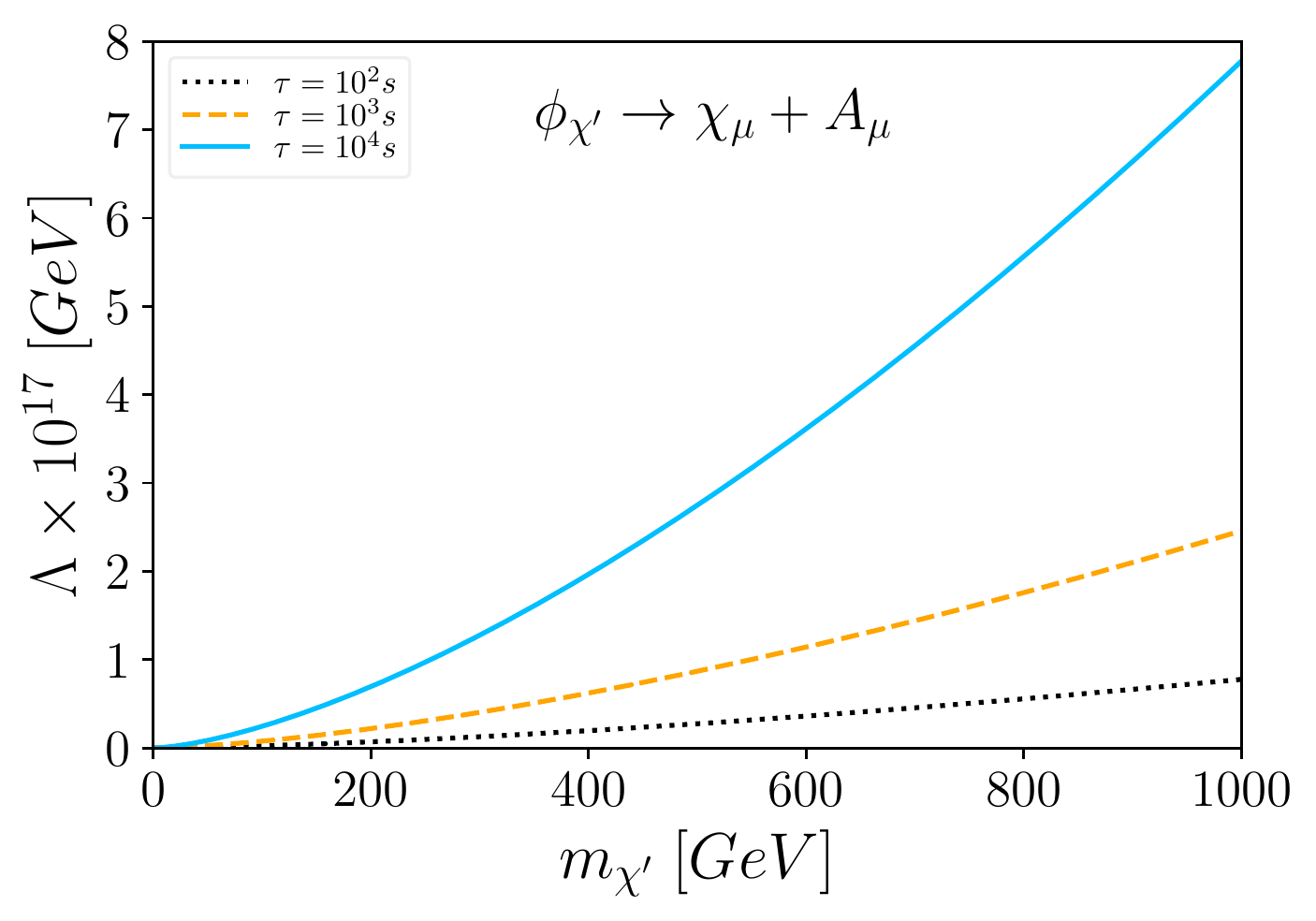}
    \label{fig:MxLambda_case3_SVG}}
    \subfigure[]{
    \includegraphics[width=\columnwidth]{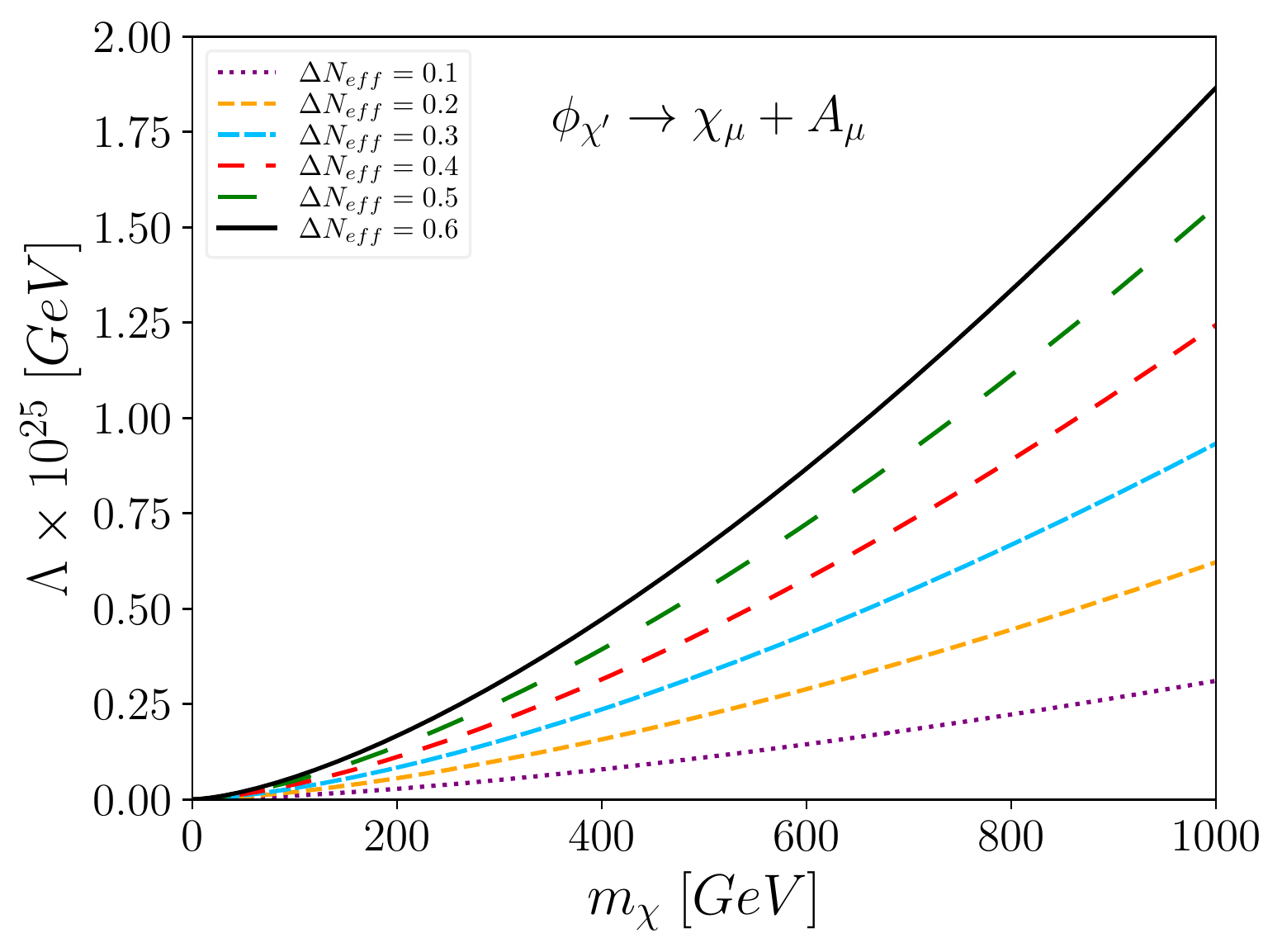}
    \label{fig:plot_Lambda_mchi_mod3}}
    \caption{Plot of $\Lambda$ as a function of the $\chi$ and $\chi^\prime$ masses for the case where $\chi^\prime$ has spin-0 and $\chi$ is a spin-1 particle. \textbf{(a)} $\Lambda \times m_{\chi^\prime}$ curves  built from Eq. (\ref{eq:gamma3_SVG}), using $\tau = 10^2, 10^3 \, \text{and} \, 10^4 \,\text{s}$. \textbf{(b)} $\Lambda \times m_{\chi}$ curves constructed from Eqs. (\ref{eq:deltaN}) and (\ref{eq:gamma3_SVG}). We consider the cases where $\Delta N_{eff} = 0.1 - 0.6$, with $f=0.01$ and $m_{\chi^\prime}/m_{\chi} = 10^4$.}
    \label{fig:plots_caseIII}
\end{figure}

The region of parameter that results in $\tau=10^2,10^3,10^4$~s are is shown in \fig{fig:MxLambda_case3_SVG}, and the parameter space that may present a solution to the $H_0$ problem is displayed in \fig{fig:plot_Lambda_mchi_mod3}, setting $m_{\chi^\prime}/m_{\chi}=10^4$, and $f=0.01$.

\section{\label{R}Results and discussions}


All the effective theories considered here feature $\tau \propto \Lambda^2/m_{\chi^\prime}^3$. Hence, the larger the mass of the mother particle the shorter the lifetime. This outcome is not new. Indeed, searches for gamma-rays and x-rays resulted from long-lived particles have been conducted exploiting \cite{Baring:2015sza}. It allows us to place a lower mass limit on its mass. To warrant a long-lived $\chi^\prime$, we need to invoke a large $\Lambda$. This explains the large energy scale, $\Lambda$, as shown in the figures \ref{fig:plots_caseI}, \ref{fig:plots_caseII} and \ref{fig:plots_caseIII}. It does not come as a surprise, because long-lived particles are related to some suppression mechanism, either present in the coupling constant or the energy scale. In our work, it is the latter \cite{Mambrini:2015sia}.


Noticed that the larger the effective energy scale, $\Lambda$ the longer the lifetime.  Consequently, larger values of $\Delta N_{eff}$ are found. Moreover, the larger $m_{\chi^\prime}$ the smaller the lifetime. However, the larger $m_{\chi^\prime}$ the larger  $\Delta N_{eff}$. Hence, there are two competing effects happening as we change $m_{\chi^\prime}$. Anyway, notice that regardless of the spin of the particles involved, non-thermally produced dark matter particles with masses at the electroweak scale can solve the $H_0$ discrepancy in agreement with BBN, CMB, and structure formation constraints. 

Interestingly, these late-decaying particles producing dark matter appear in UV complete models \cite{Feng:2003uy,Feng:2004zu,Wang:2004ib,Cembranos:2005us,Kelso:2013nwa}. An exciting outcome of our work is the correlation between particle physics, early, and late-time cosmology. 

\section{\label{dc} conclusions}

In this work, we constructed non-renomalizable operators involving spin-0, 1/2, and 1 dark matter particles produced non-thermally via the decay of a heavy companion through the process $\chi^\prime \rightarrow \chi +\gamma$. These dark matter particles are produced relativistically at the decay time. However, their energy decreases with the redshift, and they become essentially non-relativistic at matter-radiation equality for the sake of structure formation. This relativistic behavior of dark matter particles early on mimics the effect of an extra degree of freedom that helps reconcile early and late measurements of the Hubble constant. Depending on the local value adopted for the $H_0$, our mechanism might solve the $H_0$ trouble within the $\Lambda$CDM model, without evoking new dark energy densities. If $H_0$ turns out to be larger than $70 kms^{-1}Mpc^{-1}$, then indeed a new equation of state for the dark energy component is needed. Assuming that $H_0>70 kms^{-1}Mpc^{-1}$, under a Phantom-like cosmology, we showed that the typical energy scale governing the decay $\chi^\prime \rightarrow \chi +\gamma$ should range from $10^{17}$~GeV to $10^{25}$~GeV depending on the spin nature of the dark matter particle and the mass ratio $m_{\chi^\prime}/m_\chi$. Such large energy scales are natural, as one needs to invoke a large suppression mechanism to have a long-lived particle with a lifetime larger than $10^2$~s. Our work shows that perhaps the solution to the $H_0$ trouble might reside in the production mechanism of dark matter particles or a combination of both dark energy and dark matter components amenable to BBN, CMB, and structure formation observables.

\acknowledgments
ASJ acknowledges support from Coordenaç\~ao de Aperfeiçoamento de Pessoal de N\'ivel Superior (CAPES) under Grant No. 88887.497142/2020-00. D. R. da Silva thanks for the support from CAPES under grant 88887.462084/2019-00. FSQ acknowledges the Simons Foundation (Award Number:884966, AF), FAPESP grant 2021/01089-1, ICTP-SAIFR FAPESP grants 2021/14335-0, CNPq grant 307130/2021-5, and ANID- Programa Milenio - code ICN2019\_ 044. N.P.N. acknowledges the support of CNPq of Brazil under grant PQ-IB 310121/2021-3.

\bibliography{ref}%

\begin{thebibliography}{62}%
\makeatletter
\providecommand \@ifxundefined [1]{%
 \@ifx{#1\undefined}
}%
\providecommand \@ifnum [1]{%
 \ifnum #1\expandafter \@firstoftwo
 \else \expandafter \@secondoftwo
 \fi
}%
\providecommand \@ifx [1]{%
 \ifx #1\expandafter \@firstoftwo
 \else \expandafter \@secondoftwo
 \fi
}%
\providecommand \natexlab [1]{#1}%
\providecommand \enquote  [1]{``#1''}%
\providecommand \bibnamefont  [1]{#1}%
\providecommand \bibfnamefont [1]{#1}%
\providecommand \citenamefont [1]{#1}%
\providecommand \href@noop [0]{\@secondoftwo}%
\providecommand \href [0]{\begingroup \@sanitize@url \@href}%
\providecommand \@href[1]{\@@startlink{#1}\@@href}%
\providecommand \@@href[1]{\endgroup#1\@@endlink}%
\providecommand \@sanitize@url [0]{\catcode `\\12\catcode `\$12\catcode
  `\&12\catcode `\#12\catcode `\^12\catcode `\_12\catcode `\%12\relax}%
\providecommand \@@startlink[1]{}%
\providecommand \@@endlink[0]{}%
\providecommand \url  [0]{\begingroup\@sanitize@url \@url }%
\providecommand \@url [1]{\endgroup\@href {#1}{\urlprefix }}%
\providecommand \urlprefix  [0]{URL }%
\providecommand \Eprint [0]{\href }%
\providecommand \doibase [0]{http://dx.doi.org/}%
\providecommand \selectlanguage [0]{\@gobble}%
\providecommand \bibinfo  [0]{\@secondoftwo}%
\providecommand \bibfield  [0]{\@secondoftwo}%
\providecommand \translation [1]{[#1]}%
\providecommand \BibitemOpen [0]{}%
\providecommand \bibitemStop [0]{}%
\providecommand \bibitemNoStop [0]{.\EOS\space}%
\providecommand \EOS [0]{\spacefactor3000\relax}%
\providecommand \BibitemShut  [1]{\csname bibitem#1\endcsname}%
\let\auto@bib@innerbib\@empty
\bibitem [{\citenamefont {Piattella}(2018)}]{piattella2018lecture}%
  \BibitemOpen
  \bibfield  {author} {\bibinfo {author} {\bibfnamefont {O.}~\bibnamefont
  {Piattella}},\ }\href@noop {} {\emph {\bibinfo {title} {Lecture notes in
  cosmology}}}\ (\bibinfo  {publisher} {Springer},\ \bibinfo {year}
  {2018})\BibitemShut {NoStop}%
\bibitem [{\citenamefont {Hobson}\ \emph {et~al.}(2006)\citenamefont {Hobson},
  \citenamefont {Efstathiou},\ and\ \citenamefont
  {Lasenby}}]{ref:hobson2006GR}%
  \BibitemOpen
  \bibfield  {author} {\bibinfo {author} {\bibfnamefont {M.~P.}\ \bibnamefont
  {Hobson}}, \bibinfo {author} {\bibfnamefont {G.~P.}\ \bibnamefont
  {Efstathiou}}, \ and\ \bibinfo {author} {\bibfnamefont {A.~N.}\ \bibnamefont
  {Lasenby}},\ }\href@noop {} {\emph {\bibinfo {title} {General relativity: an
  introduction for physicists}}}\ (\bibinfo  {publisher} {Cambridge University
  Press},\ \bibinfo {year} {2006})\BibitemShut {NoStop}%
\bibitem [{\citenamefont {Cyburt}\ \emph {et~al.}(2016)\citenamefont {Cyburt},
  \citenamefont {Fields}, \citenamefont {Olive},\ and\ \citenamefont
  {Yeh}}]{Cyburt:2015mya}%
  \BibitemOpen
  \bibfield  {author} {\bibinfo {author} {\bibfnamefont {R.~H.}\ \bibnamefont
  {Cyburt}}, \bibinfo {author} {\bibfnamefont {B.~D.}\ \bibnamefont {Fields}},
  \bibinfo {author} {\bibfnamefont {K.~A.}\ \bibnamefont {Olive}}, \ and\
  \bibinfo {author} {\bibfnamefont {T.-H.}\ \bibnamefont {Yeh}},\ }\href
  {\doibase 10.1103/RevModPhys.88.015004} {\bibfield  {journal} {\bibinfo
  {journal} {Rev. Mod. Phys.}\ }\textbf {\bibinfo {volume} {88}},\ \bibinfo
  {pages} {015004} (\bibinfo {year} {2016})},\ \Eprint
  {http://arxiv.org/abs/1505.01076} {arXiv:1505.01076 [astro-ph.CO]}
  \BibitemShut {NoStop}%
\bibitem [{\citenamefont {Aghanim}\ \emph
  {et~al.}(2020{\natexlab{a}})\citenamefont {Aghanim} \emph
  {et~al.}}]{Planck:2018vyg}%
  \BibitemOpen
  \bibfield  {author} {\bibinfo {author} {\bibfnamefont {N.}~\bibnamefont
  {Aghanim}} \emph {et~al.} (\bibinfo {collaboration} {Planck}),\ }\href
  {\doibase 10.1051/0004-6361/201833910} {\bibfield  {journal} {\bibinfo
  {journal} {Astron. Astrophys.}\ }\textbf {\bibinfo {volume} {641}},\ \bibinfo
  {pages} {A6} (\bibinfo {year} {2020}{\natexlab{a}})},\ \bibinfo {note}
  {[Erratum: Astron.Astrophys. 652, C4 (2021)]},\ \Eprint
  {http://arxiv.org/abs/1807.06209} {arXiv:1807.06209 [astro-ph.CO]}
  \BibitemShut {NoStop}%
\bibitem [{\citenamefont {Collaboration}(2020)}]{DES:2020}%
  \BibitemOpen
  \bibfield  {author} {\bibinfo {author} {\bibfnamefont {D.~E.~S.}\
  \bibnamefont {Collaboration}},\ }\href@noop {} {\bibfield  {journal}
  {\bibinfo  {journal} {Phys. Rev. D}\ }\textbf {\bibinfo {volume} {102}},\
  \bibinfo {pages} {023509} (\bibinfo {year} {2020})},\ \Eprint
  {http://arxiv.org/abs/2002.11124} {arXiv:2002.11124 [astro-ph.CO]}
  \BibitemShut {NoStop}%
\bibitem [{\citenamefont {Blanton}\ \emph {et~al.}(2017)\citenamefont {Blanton}
  \emph {et~al.}}]{Blanton:2017}%
  \BibitemOpen
  \bibfield  {author} {\bibinfo {author} {\bibfnamefont {M.~R.}\ \bibnamefont
  {Blanton}} \emph {et~al.},\ }\href@noop {} {\bibfield  {journal} {\bibinfo
  {journal} {The Astronomical Journal}\ }\textbf {\bibinfo {volume} {154}},\
  \bibinfo {pages} {28} (\bibinfo {year} {2017})},\ \Eprint
  {http://arxiv.org/abs/1703.00052} {arXiv:1703.00052 [astro-ph.GA]}
  \BibitemShut {NoStop}%
\bibitem [{\citenamefont {Riess}\ \emph {et~al.}(1998)\citenamefont {Riess}
  \emph {et~al.}}]{Riess:1998}%
  \BibitemOpen
  \bibfield  {author} {\bibinfo {author} {\bibfnamefont {A.~G.}\ \bibnamefont
  {Riess}} \emph {et~al.},\ }\href@noop {} {\bibfield  {journal} {\bibinfo
  {journal} {The Astronomical Journal}\ }\textbf {\bibinfo {volume} {116}},\
  \bibinfo {pages} {1009} (\bibinfo {year} {1998})},\ \Eprint
  {http://arxiv.org/abs/astro-ph/9805201} {arXiv:astro-ph/9805201} \BibitemShut
  {NoStop}%
\bibitem [{\citenamefont {Perlmutter}\ \emph {et~al.}(1999)\citenamefont
  {Perlmutter} \emph {et~al.}}]{Perlmutter:1999}%
  \BibitemOpen
  \bibfield  {author} {\bibinfo {author} {\bibfnamefont {S.}~\bibnamefont
  {Perlmutter}} \emph {et~al.},\ }\href@noop {} {\bibfield  {journal} {\bibinfo
   {journal} {Astrophys.J.}\ }\textbf {\bibinfo {volume} {517}},\ \bibinfo
  {pages} {565} (\bibinfo {year} {1999})},\ \Eprint
  {http://arxiv.org/abs/astro-ph/9812133} {arXiv:astro-ph/9812133} \BibitemShut
  {NoStop}%
\bibitem [{\citenamefont {Bonvin}\ \emph {et~al.}(2016)\citenamefont {Bonvin},
  \citenamefont {Courbin}, \citenamefont {Suyu}, \citenamefont {Marshall},
  \citenamefont {Rusu}, \citenamefont {Sluse}, \citenamefont {Tewes},
  \citenamefont {Wong}, \citenamefont {Collett}, \citenamefont {Fassnacht},
  \citenamefont {Treu}, \citenamefont {Auger}, \citenamefont {Hilbert},
  \citenamefont {Koopmans}, \citenamefont {Meylan}, \citenamefont {Rumbaugh},
  \citenamefont {Sonnenfeld},\ and\ \citenamefont {Spiniello}}]{H0LiCOW2016}%
  \BibitemOpen
  \bibfield  {author} {\bibinfo {author} {\bibfnamefont {V.}~\bibnamefont
  {Bonvin}}, \bibinfo {author} {\bibfnamefont {F.}~\bibnamefont {Courbin}},
  \bibinfo {author} {\bibfnamefont {S.~H.}\ \bibnamefont {Suyu}}, \bibinfo
  {author} {\bibfnamefont {P.~J.}\ \bibnamefont {Marshall}}, \bibinfo {author}
  {\bibfnamefont {C.~E.}\ \bibnamefont {Rusu}}, \bibinfo {author}
  {\bibfnamefont {D.}~\bibnamefont {Sluse}}, \bibinfo {author} {\bibfnamefont
  {M.}~\bibnamefont {Tewes}}, \bibinfo {author} {\bibfnamefont {K.~C.}\
  \bibnamefont {Wong}}, \bibinfo {author} {\bibfnamefont {T.}~\bibnamefont
  {Collett}}, \bibinfo {author} {\bibfnamefont {C.~D.}\ \bibnamefont
  {Fassnacht}}, \bibinfo {author} {\bibfnamefont {T.}~\bibnamefont {Treu}},
  \bibinfo {author} {\bibfnamefont {M.~W.}\ \bibnamefont {Auger}}, \bibinfo
  {author} {\bibfnamefont {S.}~\bibnamefont {Hilbert}}, \bibinfo {author}
  {\bibfnamefont {L.~V.~E.}\ \bibnamefont {Koopmans}}, \bibinfo {author}
  {\bibfnamefont {G.}~\bibnamefont {Meylan}}, \bibinfo {author} {\bibfnamefont
  {N.}~\bibnamefont {Rumbaugh}}, \bibinfo {author} {\bibfnamefont
  {A.}~\bibnamefont {Sonnenfeld}}, \ and\ \bibinfo {author} {\bibfnamefont
  {C.}~\bibnamefont {Spiniello}},\ }\href {\doibase 10.1093/mnras/stw3006}
  {\bibfield  {journal} {\bibinfo  {journal} {Monthly Notices of the Royal
  Astronomical Society}\ }\textbf {\bibinfo {volume} {465}},\ \bibinfo {pages}
  {4914} (\bibinfo {year} {2016})},\ \Eprint
  {http://arxiv.org/abs/https://academic.oup.com/mnras/article-pdf/465/4/4914/10254866/stw3006.pdf}
  {https://academic.oup.com/mnras/article-pdf/465/4/4914/10254866/stw3006.pdf}
  \BibitemShut {NoStop}%
\bibitem [{\citenamefont {Bernal}\ \emph {et~al.}(2016)\citenamefont {Bernal},
  \citenamefont {Verde},\ and\ \citenamefont {Riess}}]{ref:Bernal2016}%
  \BibitemOpen
  \bibfield  {author} {\bibinfo {author} {\bibfnamefont {J.~L.}\ \bibnamefont
  {Bernal}}, \bibinfo {author} {\bibfnamefont {L.}~\bibnamefont {Verde}}, \
  and\ \bibinfo {author} {\bibfnamefont {A.~G.}\ \bibnamefont {Riess}},\ }\href
  {\doibase 10.1088/1475-7516/2016/10/019} {\bibfield  {journal} {\bibinfo
  {journal} {JCAP}\ }\textbf {\bibinfo {volume} {10}},\ \bibinfo {pages} {019}
  (\bibinfo {year} {2016})},\ \Eprint {http://arxiv.org/abs/1607.05617}
  {arXiv:1607.05617 [astro-ph.CO]} \BibitemShut {NoStop}%
\bibitem [{\citenamefont {Anchordoqui}\ \emph {et~al.}(2021)\citenamefont
  {Anchordoqui}, \citenamefont {Di~Valentino}, \citenamefont {Pan},\ and\
  \citenamefont {Yang}}]{ref:Anchordoqui2021}%
  \BibitemOpen
  \bibfield  {author} {\bibinfo {author} {\bibfnamefont {L.~A.}\ \bibnamefont
  {Anchordoqui}}, \bibinfo {author} {\bibfnamefont {E.}~\bibnamefont
  {Di~Valentino}}, \bibinfo {author} {\bibfnamefont {S.}~\bibnamefont {Pan}}, \
  and\ \bibinfo {author} {\bibfnamefont {W.}~\bibnamefont {Yang}},\ }\href
  {\doibase 10.1016/j.jheap.2021.08.001} {\bibfield  {journal} {\bibinfo
  {journal} {JHEAp}\ }\textbf {\bibinfo {volume} {32}},\ \bibinfo {pages} {28}
  (\bibinfo {year} {2021})},\ \Eprint {http://arxiv.org/abs/2107.13932}
  {arXiv:2107.13932 [astro-ph.CO]} \BibitemShut {NoStop}%
\bibitem [{\citenamefont {Hinshaw}\ \emph {et~al.}(2013)\citenamefont
  {Hinshaw}, \citenamefont {Larson}, \citenamefont {Komatsu}, \citenamefont
  {Spergel}, \citenamefont {Bennett}, \citenamefont {Dunkley}, \citenamefont
  {Nolta}, \citenamefont {Halpern}, \citenamefont {Hill}, \citenamefont
  {Odegard}, \citenamefont {Page}, \citenamefont {Smith}, \citenamefont
  {Weiland}, \citenamefont {Gold}, \citenamefont {Jarosik}, \citenamefont
  {Kogut}, \citenamefont {Limon}, \citenamefont {Meyer}, \citenamefont
  {Tucker}, \citenamefont {Wollack},\ and\ \citenamefont
  {Wright}}]{Hinshaw_2013}%
  \BibitemOpen
  \bibfield  {author} {\bibinfo {author} {\bibfnamefont {G.}~\bibnamefont
  {Hinshaw}}, \bibinfo {author} {\bibfnamefont {D.}~\bibnamefont {Larson}},
  \bibinfo {author} {\bibfnamefont {E.}~\bibnamefont {Komatsu}}, \bibinfo
  {author} {\bibfnamefont {D.~N.}\ \bibnamefont {Spergel}}, \bibinfo {author}
  {\bibfnamefont {C.~L.}\ \bibnamefont {Bennett}}, \bibinfo {author}
  {\bibfnamefont {J.}~\bibnamefont {Dunkley}}, \bibinfo {author} {\bibfnamefont
  {M.~R.}\ \bibnamefont {Nolta}}, \bibinfo {author} {\bibfnamefont
  {M.}~\bibnamefont {Halpern}}, \bibinfo {author} {\bibfnamefont {R.~S.}\
  \bibnamefont {Hill}}, \bibinfo {author} {\bibfnamefont {N.}~\bibnamefont
  {Odegard}}, \bibinfo {author} {\bibfnamefont {L.}~\bibnamefont {Page}},
  \bibinfo {author} {\bibfnamefont {K.~M.}\ \bibnamefont {Smith}}, \bibinfo
  {author} {\bibfnamefont {J.~L.}\ \bibnamefont {Weiland}}, \bibinfo {author}
  {\bibfnamefont {B.}~\bibnamefont {Gold}}, \bibinfo {author} {\bibfnamefont
  {N.}~\bibnamefont {Jarosik}}, \bibinfo {author} {\bibfnamefont
  {A.}~\bibnamefont {Kogut}}, \bibinfo {author} {\bibfnamefont
  {M.}~\bibnamefont {Limon}}, \bibinfo {author} {\bibfnamefont {S.~S.}\
  \bibnamefont {Meyer}}, \bibinfo {author} {\bibfnamefont {G.~S.}\ \bibnamefont
  {Tucker}}, \bibinfo {author} {\bibfnamefont {E.}~\bibnamefont {Wollack}}, \
  and\ \bibinfo {author} {\bibfnamefont {E.~L.}\ \bibnamefont {Wright}},\
  }\href {\doibase 10.1088/0067-0049/208/2/19} {\bibfield  {journal} {\bibinfo
  {journal} {The Astrophysical Journal Supplement Series}\ }\textbf {\bibinfo
  {volume} {208}},\ \bibinfo {pages} {19} (\bibinfo {year} {2013})}\BibitemShut
  {NoStop}%
\bibitem [{\citenamefont {Dutcher}\ \emph {et~al.}(2021)\citenamefont {Dutcher}
  \emph {et~al.}}]{SPT-3G:2021eoc}%
  \BibitemOpen
  \bibfield  {author} {\bibinfo {author} {\bibfnamefont {D.}~\bibnamefont
  {Dutcher}} \emph {et~al.} (\bibinfo {collaboration} {SPT-3G}),\ }\href
  {\doibase 10.1103/PhysRevD.104.022003} {\bibfield  {journal} {\bibinfo
  {journal} {Phys. Rev. D}\ }\textbf {\bibinfo {volume} {104}},\ \bibinfo
  {pages} {022003} (\bibinfo {year} {2021})},\ \Eprint
  {http://arxiv.org/abs/2101.01684} {arXiv:2101.01684 [astro-ph.CO]}
  \BibitemShut {NoStop}%
\bibitem [{\citenamefont {Wang}\ and\ \citenamefont
  {Huang}(2020)}]{Wang:2019isw}%
  \BibitemOpen
  \bibfield  {author} {\bibinfo {author} {\bibfnamefont {K.}~\bibnamefont
  {Wang}}\ and\ \bibinfo {author} {\bibfnamefont {Q.-G.}\ \bibnamefont
  {Huang}},\ }\href {\doibase 10.1088/1475-7516/2020/06/045} {\bibfield
  {journal} {\bibinfo  {journal} {JCAP}\ }\textbf {\bibinfo {volume} {06}},\
  \bibinfo {pages} {045} (\bibinfo {year} {2020})},\ \Eprint
  {http://arxiv.org/abs/1912.05491} {arXiv:1912.05491 [astro-ph.CO]}
  \BibitemShut {NoStop}%
\bibitem [{\citenamefont {Ivanov}\ \emph {et~al.}(2020)\citenamefont {Ivanov},
  \citenamefont {Simonovi\'c},\ and\ \citenamefont
  {Zaldarriaga}}]{Ivanov:2019pdj}%
  \BibitemOpen
  \bibfield  {author} {\bibinfo {author} {\bibfnamefont {M.~M.}\ \bibnamefont
  {Ivanov}}, \bibinfo {author} {\bibfnamefont {M.}~\bibnamefont {Simonovi\'c}},
  \ and\ \bibinfo {author} {\bibfnamefont {M.}~\bibnamefont {Zaldarriaga}},\
  }\href {\doibase 10.1088/1475-7516/2020/05/042} {\bibfield  {journal}
  {\bibinfo  {journal} {JCAP}\ }\textbf {\bibinfo {volume} {05}},\ \bibinfo
  {pages} {042} (\bibinfo {year} {2020})},\ \Eprint
  {http://arxiv.org/abs/1909.05277} {arXiv:1909.05277 [astro-ph.CO]}
  \BibitemShut {NoStop}%
\bibitem [{\citenamefont {Alam}\ \emph {et~al.}(2021)\citenamefont {Alam} \emph
  {et~al.}}]{eBOSS:2020yzd}%
  \BibitemOpen
  \bibfield  {author} {\bibinfo {author} {\bibfnamefont {S.}~\bibnamefont
  {Alam}} \emph {et~al.} (\bibinfo {collaboration} {eBOSS}),\ }\href {\doibase
  10.1103/PhysRevD.103.083533} {\bibfield  {journal} {\bibinfo  {journal}
  {Phys. Rev. D}\ }\textbf {\bibinfo {volume} {103}},\ \bibinfo {pages}
  {083533} (\bibinfo {year} {2021})},\ \Eprint
  {http://arxiv.org/abs/2007.08991} {arXiv:2007.08991 [astro-ph.CO]}
  \BibitemShut {NoStop}%
\bibitem [{\citenamefont {Cardona}\ \emph {et~al.}(2017)\citenamefont
  {Cardona}, \citenamefont {Kunz},\ and\ \citenamefont
  {Pettorino}}]{Cardona:2016ems}%
  \BibitemOpen
  \bibfield  {author} {\bibinfo {author} {\bibfnamefont {W.}~\bibnamefont
  {Cardona}}, \bibinfo {author} {\bibfnamefont {M.}~\bibnamefont {Kunz}}, \
  and\ \bibinfo {author} {\bibfnamefont {V.}~\bibnamefont {Pettorino}},\ }\href
  {\doibase 10.1088/1475-7516/2017/03/056} {\bibfield  {journal} {\bibinfo
  {journal} {JCAP}\ }\textbf {\bibinfo {volume} {03}},\ \bibinfo {pages} {056}
  (\bibinfo {year} {2017})},\ \Eprint {http://arxiv.org/abs/1611.06088}
  {arXiv:1611.06088 [astro-ph.CO]} \BibitemShut {NoStop}%
\bibitem [{\citenamefont {Camarena}\ and\ \citenamefont
  {Marra}(2020)}]{Camarena:2019moy}%
  \BibitemOpen
  \bibfield  {author} {\bibinfo {author} {\bibfnamefont {D.}~\bibnamefont
  {Camarena}}\ and\ \bibinfo {author} {\bibfnamefont {V.}~\bibnamefont
  {Marra}},\ }\href {\doibase 10.1103/PhysRevResearch.2.013028} {\bibfield
  {journal} {\bibinfo  {journal} {Phys. Rev. Res.}\ }\textbf {\bibinfo {volume}
  {2}},\ \bibinfo {pages} {013028} (\bibinfo {year} {2020})},\ \Eprint
  {http://arxiv.org/abs/1906.11814} {arXiv:1906.11814 [astro-ph.CO]}
  \BibitemShut {NoStop}%
\bibitem [{\citenamefont {Breuval}\ \emph {et~al.}(2020)\citenamefont {Breuval}
  \emph {et~al.}}]{Breuval:2020trd}%
  \BibitemOpen
  \bibfield  {author} {\bibinfo {author} {\bibfnamefont {L.}~\bibnamefont
  {Breuval}} \emph {et~al.},\ }\href {\doibase 10.1051/0004-6361/202038633}
  {\bibfield  {journal} {\bibinfo  {journal} {Astron. Astrophys.}\ }\textbf
  {\bibinfo {volume} {643}},\ \bibinfo {pages} {A115} (\bibinfo {year}
  {2020})},\ \Eprint {http://arxiv.org/abs/2006.08763} {arXiv:2006.08763
  [astro-ph.SR]} \BibitemShut {NoStop}%
\bibitem [{\citenamefont {Riess}\ \emph {et~al.}(2021)\citenamefont {Riess},
  \citenamefont {Casertano}, \citenamefont {Yuan}, \citenamefont {Bowers},
  \citenamefont {Macri}, \citenamefont {Zinn},\ and\ \citenamefont
  {Scolnic}}]{Riess:2020fzl}%
  \BibitemOpen
  \bibfield  {author} {\bibinfo {author} {\bibfnamefont {A.~G.}\ \bibnamefont
  {Riess}}, \bibinfo {author} {\bibfnamefont {S.}~\bibnamefont {Casertano}},
  \bibinfo {author} {\bibfnamefont {W.}~\bibnamefont {Yuan}}, \bibinfo {author}
  {\bibfnamefont {J.~B.}\ \bibnamefont {Bowers}}, \bibinfo {author}
  {\bibfnamefont {L.}~\bibnamefont {Macri}}, \bibinfo {author} {\bibfnamefont
  {J.~C.}\ \bibnamefont {Zinn}}, \ and\ \bibinfo {author} {\bibfnamefont
  {D.}~\bibnamefont {Scolnic}},\ }\href {\doibase 10.3847/2041-8213/abdbaf}
  {\bibfield  {journal} {\bibinfo  {journal} {Astrophys. J. Lett.}\ }\textbf
  {\bibinfo {volume} {908}},\ \bibinfo {pages} {L6} (\bibinfo {year} {2021})},\
  \Eprint {http://arxiv.org/abs/2012.08534} {arXiv:2012.08534 [astro-ph.CO]}
  \BibitemShut {NoStop}%
\bibitem [{\citenamefont {Freedman}\ \emph {et~al.}(2019)\citenamefont
  {Freedman} \emph {et~al.}}]{Freedman:2019jwv}%
  \BibitemOpen
  \bibfield  {author} {\bibinfo {author} {\bibfnamefont {W.~L.}\ \bibnamefont
  {Freedman}} \emph {et~al.},\ }\href {\doibase 10.3847/1538-4357/ab2f73} {\
  (\bibinfo {year} {2019}),\ 10.3847/1538-4357/ab2f73},\ \Eprint
  {http://arxiv.org/abs/1907.05922} {arXiv:1907.05922 [astro-ph.CO]}
  \BibitemShut {NoStop}%
\bibitem [{\citenamefont {Calabrese}\ \emph {et~al.}(2008)\citenamefont
  {Calabrese}, \citenamefont {Slosar}, \citenamefont {Melchiorri},
  \citenamefont {Smoot},\ and\ \citenamefont {Zahn}}]{Calabrese:2008rt}%
  \BibitemOpen
  \bibfield  {author} {\bibinfo {author} {\bibfnamefont {E.}~\bibnamefont
  {Calabrese}}, \bibinfo {author} {\bibfnamefont {A.}~\bibnamefont {Slosar}},
  \bibinfo {author} {\bibfnamefont {A.}~\bibnamefont {Melchiorri}}, \bibinfo
  {author} {\bibfnamefont {G.~F.}\ \bibnamefont {Smoot}}, \ and\ \bibinfo
  {author} {\bibfnamefont {O.}~\bibnamefont {Zahn}},\ }\href {\doibase
  10.1103/PhysRevD.77.123531} {\bibfield  {journal} {\bibinfo  {journal} {Phys.
  Rev. D}\ }\textbf {\bibinfo {volume} {77}},\ \bibinfo {pages} {123531}
  (\bibinfo {year} {2008})},\ \Eprint {http://arxiv.org/abs/0803.2309}
  {arXiv:0803.2309 [astro-ph]} \BibitemShut {NoStop}%
\bibitem [{\citenamefont {Freedman}(2021)}]{Freedman:2021ahq}%
  \BibitemOpen
  \bibfield  {author} {\bibinfo {author} {\bibfnamefont {W.~L.}\ \bibnamefont
  {Freedman}},\ }\href {\doibase 10.3847/1538-4357/ac0e95} {\bibfield
  {journal} {\bibinfo  {journal} {Astrophys. J.}\ }\textbf {\bibinfo {volume}
  {919}},\ \bibinfo {pages} {16} (\bibinfo {year} {2021})},\ \Eprint
  {http://arxiv.org/abs/2106.15656} {arXiv:2106.15656 [astro-ph.CO]}
  \BibitemShut {NoStop}%
\bibitem [{\citenamefont {Di~Valentino}\ \emph {et~al.}(2021)\citenamefont
  {Di~Valentino}, \citenamefont {Mena}, \citenamefont {Pan}, \citenamefont
  {Visinelli}, \citenamefont {Yang}, \citenamefont {Melchiorri}, \citenamefont
  {Mota}, \citenamefont {Riess},\ and\ \citenamefont
  {Silk}}]{DiValentino:2021izs}%
  \BibitemOpen
  \bibfield  {author} {\bibinfo {author} {\bibfnamefont {E.}~\bibnamefont
  {Di~Valentino}}, \bibinfo {author} {\bibfnamefont {O.}~\bibnamefont {Mena}},
  \bibinfo {author} {\bibfnamefont {S.}~\bibnamefont {Pan}}, \bibinfo {author}
  {\bibfnamefont {L.}~\bibnamefont {Visinelli}}, \bibinfo {author}
  {\bibfnamefont {W.}~\bibnamefont {Yang}}, \bibinfo {author} {\bibfnamefont
  {A.}~\bibnamefont {Melchiorri}}, \bibinfo {author} {\bibfnamefont {D.~F.}\
  \bibnamefont {Mota}}, \bibinfo {author} {\bibfnamefont {A.~G.}\ \bibnamefont
  {Riess}}, \ and\ \bibinfo {author} {\bibfnamefont {J.}~\bibnamefont {Silk}},\
  }\href {\doibase 10.1088/1361-6382/ac086d} {\bibfield  {journal} {\bibinfo
  {journal} {Class. Quant. Grav.}\ }\textbf {\bibinfo {volume} {38}},\ \bibinfo
  {pages} {153001} (\bibinfo {year} {2021})},\ \Eprint
  {http://arxiv.org/abs/2103.01183} {arXiv:2103.01183 [astro-ph.CO]}
  \BibitemShut {NoStop}%
\bibitem [{\citenamefont {Kamionkowski}\ and\ \citenamefont
  {Riess}(2022)}]{Kamionkowski:2022pkx}%
  \BibitemOpen
  \bibfield  {author} {\bibinfo {author} {\bibfnamefont {M.}~\bibnamefont
  {Kamionkowski}}\ and\ \bibinfo {author} {\bibfnamefont {A.~G.}\ \bibnamefont
  {Riess}},\ }\href@noop {} {\  (\bibinfo {year} {2022})},\ \Eprint
  {http://arxiv.org/abs/2211.04492} {arXiv:2211.04492 [astro-ph.CO]}
  \BibitemShut {NoStop}%
\bibitem [{\citenamefont {Anchordoqui}\ and\ \citenamefont
  {Goldberg}(2012)}]{Anchordoqui:2011nh}%
  \BibitemOpen
  \bibfield  {author} {\bibinfo {author} {\bibfnamefont {L.~A.}\ \bibnamefont
  {Anchordoqui}}\ and\ \bibinfo {author} {\bibfnamefont {H.}~\bibnamefont
  {Goldberg}},\ }\href {\doibase 10.1103/PhysRevLett.108.081805} {\bibfield
  {journal} {\bibinfo  {journal} {Phys. Rev. Lett.}\ }\textbf {\bibinfo
  {volume} {108}},\ \bibinfo {pages} {081805} (\bibinfo {year} {2012})},\
  \Eprint {http://arxiv.org/abs/1111.7264} {arXiv:1111.7264 [hep-ph]}
  \BibitemShut {NoStop}%
\bibitem [{\citenamefont {Boehm}\ \emph {et~al.}(2012)\citenamefont {Boehm},
  \citenamefont {Dolan},\ and\ \citenamefont {McCabe}}]{Boehm:2012gr}%
  \BibitemOpen
  \bibfield  {author} {\bibinfo {author} {\bibfnamefont {C.}~\bibnamefont
  {Boehm}}, \bibinfo {author} {\bibfnamefont {M.~J.}\ \bibnamefont {Dolan}}, \
  and\ \bibinfo {author} {\bibfnamefont {C.}~\bibnamefont {McCabe}},\ }\href
  {\doibase 10.1088/1475-7516/2012/12/027} {\bibfield  {journal} {\bibinfo
  {journal} {JCAP}\ }\textbf {\bibinfo {volume} {12}},\ \bibinfo {pages} {027}
  (\bibinfo {year} {2012})},\ \Eprint {http://arxiv.org/abs/1207.0497}
  {arXiv:1207.0497 [astro-ph.CO]} \BibitemShut {NoStop}%
\bibitem [{\citenamefont {Brust}\ \emph {et~al.}(2013)\citenamefont {Brust},
  \citenamefont {Kaplan},\ and\ \citenamefont {Walters}}]{Brust:2013ova}%
  \BibitemOpen
  \bibfield  {author} {\bibinfo {author} {\bibfnamefont {C.}~\bibnamefont
  {Brust}}, \bibinfo {author} {\bibfnamefont {D.~E.}\ \bibnamefont {Kaplan}}, \
  and\ \bibinfo {author} {\bibfnamefont {M.~T.}\ \bibnamefont {Walters}},\
  }\href {\doibase 10.1007/JHEP12(2013)058} {\bibfield  {journal} {\bibinfo
  {journal} {JHEP}\ }\textbf {\bibinfo {volume} {12}},\ \bibinfo {pages} {058}
  (\bibinfo {year} {2013})},\ \Eprint {http://arxiv.org/abs/1303.5379}
  {arXiv:1303.5379 [hep-ph]} \BibitemShut {NoStop}%
\bibitem [{\citenamefont {Hooper}\ \emph {et~al.}(2012)\citenamefont {Hooper},
  \citenamefont {Queiroz},\ and\ \citenamefont {Gnedin}}]{Hooper:2011aj}%
  \BibitemOpen
  \bibfield  {author} {\bibinfo {author} {\bibfnamefont {D.}~\bibnamefont
  {Hooper}}, \bibinfo {author} {\bibfnamefont {F.~S.}\ \bibnamefont {Queiroz}},
  \ and\ \bibinfo {author} {\bibfnamefont {N.~Y.}\ \bibnamefont {Gnedin}},\
  }\href {\doibase 10.1103/PhysRevD.85.063513} {\bibfield  {journal} {\bibinfo
  {journal} {Phys. Rev. D}\ }\textbf {\bibinfo {volume} {85}},\ \bibinfo
  {pages} {063513} (\bibinfo {year} {2012})},\ \Eprint
  {http://arxiv.org/abs/1111.6599} {arXiv:1111.6599 [astro-ph.CO]} \BibitemShut
  {NoStop}%
\bibitem [{\citenamefont {Bringmann}\ \emph {et~al.}(2018)\citenamefont
  {Bringmann}, \citenamefont {Kahlhoefer}, \citenamefont {Schmidt-Hoberg},\
  and\ \citenamefont {Walia}}]{Bringmann:2018jpr}%
  \BibitemOpen
  \bibfield  {author} {\bibinfo {author} {\bibfnamefont {T.}~\bibnamefont
  {Bringmann}}, \bibinfo {author} {\bibfnamefont {F.}~\bibnamefont
  {Kahlhoefer}}, \bibinfo {author} {\bibfnamefont {K.}~\bibnamefont
  {Schmidt-Hoberg}}, \ and\ \bibinfo {author} {\bibfnamefont {P.}~\bibnamefont
  {Walia}},\ }\href {\doibase 10.1103/PhysRevD.98.023543} {\bibfield  {journal}
  {\bibinfo  {journal} {Phys. Rev. D}\ }\textbf {\bibinfo {volume} {98}},\
  \bibinfo {pages} {023543} (\bibinfo {year} {2018})},\ \Eprint
  {http://arxiv.org/abs/1803.03644} {arXiv:1803.03644 [astro-ph.CO]}
  \BibitemShut {NoStop}%
\bibitem [{\citenamefont {Caldwell}(2002)}]{Caldwell:1999ew}%
  \BibitemOpen
  \bibfield  {author} {\bibinfo {author} {\bibfnamefont {R.~R.}\ \bibnamefont
  {Caldwell}},\ }\href {\doibase 10.1016/S0370-2693(02)02589-3} {\bibfield
  {journal} {\bibinfo  {journal} {Phys. Lett. B}\ }\textbf {\bibinfo {volume}
  {545}},\ \bibinfo {pages} {23} (\bibinfo {year} {2002})},\ \Eprint
  {http://arxiv.org/abs/astro-ph/9908168} {arXiv:astro-ph/9908168} \BibitemShut
  {NoStop}%
\bibitem [{\citenamefont {Caldwell}\ \emph {et~al.}(2003)\citenamefont
  {Caldwell}, \citenamefont {Kamionkowski},\ and\ \citenamefont
  {Weinberg}}]{Caldwell:2003vq}%
  \BibitemOpen
  \bibfield  {author} {\bibinfo {author} {\bibfnamefont {R.~R.}\ \bibnamefont
  {Caldwell}}, \bibinfo {author} {\bibfnamefont {M.}~\bibnamefont
  {Kamionkowski}}, \ and\ \bibinfo {author} {\bibfnamefont {N.~N.}\
  \bibnamefont {Weinberg}},\ }\href {\doibase 10.1103/PhysRevLett.91.071301}
  {\bibfield  {journal} {\bibinfo  {journal} {Phys. Rev. Lett.}\ }\textbf
  {\bibinfo {volume} {91}},\ \bibinfo {pages} {071301} (\bibinfo {year}
  {2003})},\ \Eprint {http://arxiv.org/abs/astro-ph/0302506}
  {arXiv:astro-ph/0302506} \BibitemShut {NoStop}%
\bibitem [{\citenamefont {Vikman}(2005)}]{Vikman:2004dc}%
  \BibitemOpen
  \bibfield  {author} {\bibinfo {author} {\bibfnamefont {A.}~\bibnamefont
  {Vikman}},\ }\href {\doibase 10.1103/PhysRevD.71.023515} {\bibfield
  {journal} {\bibinfo  {journal} {Phys. Rev. D}\ }\textbf {\bibinfo {volume}
  {71}},\ \bibinfo {pages} {023515} (\bibinfo {year} {2005})},\ \Eprint
  {http://arxiv.org/abs/astro-ph/0407107} {arXiv:astro-ph/0407107} \BibitemShut
  {NoStop}%
\bibitem [{\citenamefont {Nojiri}\ \emph {et~al.}(2005)\citenamefont {Nojiri},
  \citenamefont {Odintsov},\ and\ \citenamefont {Tsujikawa}}]{Nojiri:2005sx}%
  \BibitemOpen
  \bibfield  {author} {\bibinfo {author} {\bibfnamefont {S.}~\bibnamefont
  {Nojiri}}, \bibinfo {author} {\bibfnamefont {S.~D.}\ \bibnamefont
  {Odintsov}}, \ and\ \bibinfo {author} {\bibfnamefont {S.}~\bibnamefont
  {Tsujikawa}},\ }\href {\doibase 10.1103/PhysRevD.71.063004} {\bibfield
  {journal} {\bibinfo  {journal} {Phys. Rev. D}\ }\textbf {\bibinfo {volume}
  {71}},\ \bibinfo {pages} {063004} (\bibinfo {year} {2005})},\ \Eprint
  {http://arxiv.org/abs/hep-th/0501025} {arXiv:hep-th/0501025} \BibitemShut
  {NoStop}%
\bibitem [{\citenamefont {Shafer}\ and\ \citenamefont
  {Huterer}(2014)}]{Shafer:2013pxa}%
  \BibitemOpen
  \bibfield  {author} {\bibinfo {author} {\bibfnamefont {D.~L.}\ \bibnamefont
  {Shafer}}\ and\ \bibinfo {author} {\bibfnamefont {D.}~\bibnamefont
  {Huterer}},\ }\href {\doibase 10.1103/PhysRevD.89.063510} {\bibfield
  {journal} {\bibinfo  {journal} {Phys. Rev. D}\ }\textbf {\bibinfo {volume}
  {89}},\ \bibinfo {pages} {063510} (\bibinfo {year} {2014})},\ \Eprint
  {http://arxiv.org/abs/1312.1688} {arXiv:1312.1688 [astro-ph.CO]} \BibitemShut
  {NoStop}%
\bibitem [{\citenamefont {Ludwick}(2017)}]{Ludwick:2017tox}%
  \BibitemOpen
  \bibfield  {author} {\bibinfo {author} {\bibfnamefont {K.~J.}\ \bibnamefont
  {Ludwick}},\ }\href {\doibase 10.1142/S0217732317300257} {\bibfield
  {journal} {\bibinfo  {journal} {Mod. Phys. Lett. A}\ }\textbf {\bibinfo
  {volume} {32}},\ \bibinfo {pages} {1730025} (\bibinfo {year} {2017})},\
  \Eprint {http://arxiv.org/abs/1708.06981} {arXiv:1708.06981 [astro-ph.CO]}
  \BibitemShut {NoStop}%
\bibitem [{\citenamefont {Alestas}\ \emph {et~al.}(2020)\citenamefont
  {Alestas}, \citenamefont {Kazantzidis},\ and\ \citenamefont
  {Perivolaropoulos}}]{Alestas:2020mvb}%
  \BibitemOpen
  \bibfield  {author} {\bibinfo {author} {\bibfnamefont {G.}~\bibnamefont
  {Alestas}}, \bibinfo {author} {\bibfnamefont {L.}~\bibnamefont
  {Kazantzidis}}, \ and\ \bibinfo {author} {\bibfnamefont {L.}~\bibnamefont
  {Perivolaropoulos}},\ }\href {\doibase 10.1103/PhysRevD.101.123516}
  {\bibfield  {journal} {\bibinfo  {journal} {Phys. Rev. D}\ }\textbf {\bibinfo
  {volume} {101}},\ \bibinfo {pages} {123516} (\bibinfo {year} {2020})},\
  \Eprint {http://arxiv.org/abs/2004.08363} {arXiv:2004.08363 [astro-ph.CO]}
  \BibitemShut {NoStop}%
\bibitem [{\citenamefont {Riess}\ \emph {et~al.}(2022)\citenamefont {Riess}
  \emph {et~al.}}]{Riess:2021jrx}%
  \BibitemOpen
  \bibfield  {author} {\bibinfo {author} {\bibfnamefont {A.~G.}\ \bibnamefont
  {Riess}} \emph {et~al.},\ }\href {\doibase 10.3847/2041-8213/ac5c5b}
  {\bibfield  {journal} {\bibinfo  {journal} {Astrophys. J. Lett.}\ }\textbf
  {\bibinfo {volume} {934}},\ \bibinfo {pages} {L7} (\bibinfo {year} {2022})},\
  \Eprint {http://arxiv.org/abs/2112.04510} {arXiv:2112.04510 [astro-ph.CO]}
  \BibitemShut {NoStop}%
\bibitem [{\citenamefont {Benevento}\ \emph {et~al.}(2020)\citenamefont
  {Benevento}, \citenamefont {Hu},\ and\ \citenamefont
  {Raveri}}]{Benevento:2020fev}%
  \BibitemOpen
  \bibfield  {author} {\bibinfo {author} {\bibfnamefont {G.}~\bibnamefont
  {Benevento}}, \bibinfo {author} {\bibfnamefont {W.}~\bibnamefont {Hu}}, \
  and\ \bibinfo {author} {\bibfnamefont {M.}~\bibnamefont {Raveri}},\ }\href
  {\doibase 10.1103/PhysRevD.101.103517} {\bibfield  {journal} {\bibinfo
  {journal} {Phys. Rev. D}\ }\textbf {\bibinfo {volume} {101}},\ \bibinfo
  {pages} {103517} (\bibinfo {year} {2020})},\ \Eprint
  {http://arxiv.org/abs/2002.11707} {arXiv:2002.11707 [astro-ph.CO]}
  \BibitemShut {NoStop}%
\bibitem [{\citenamefont {Alestas}\ \emph {et~al.}(2021)\citenamefont
  {Alestas}, \citenamefont {Kazantzidis},\ and\ \citenamefont
  {Perivolaropoulos}}]{Alestas:2020zol}%
  \BibitemOpen
  \bibfield  {author} {\bibinfo {author} {\bibfnamefont {G.}~\bibnamefont
  {Alestas}}, \bibinfo {author} {\bibfnamefont {L.}~\bibnamefont
  {Kazantzidis}}, \ and\ \bibinfo {author} {\bibfnamefont {L.}~\bibnamefont
  {Perivolaropoulos}},\ }\href {\doibase 10.1103/PhysRevD.103.083517}
  {\bibfield  {journal} {\bibinfo  {journal} {Phys. Rev. D}\ }\textbf {\bibinfo
  {volume} {103}},\ \bibinfo {pages} {083517} (\bibinfo {year} {2021})},\
  \Eprint {http://arxiv.org/abs/2012.13932} {arXiv:2012.13932 [astro-ph.CO]}
  \BibitemShut {NoStop}%
\bibitem [{\citenamefont {Camarena}\ and\ \citenamefont
  {Marra}(2021)}]{Camarena:2021jlr}%
  \BibitemOpen
  \bibfield  {author} {\bibinfo {author} {\bibfnamefont {D.}~\bibnamefont
  {Camarena}}\ and\ \bibinfo {author} {\bibfnamefont {V.}~\bibnamefont
  {Marra}},\ }\href {\doibase 10.1093/mnras/stab1200} {\bibfield  {journal}
  {\bibinfo  {journal} {Mon. Not. Roy. Astron. Soc.}\ }\textbf {\bibinfo
  {volume} {504}},\ \bibinfo {pages} {5164} (\bibinfo {year} {2021})},\ \Eprint
  {http://arxiv.org/abs/2101.08641} {arXiv:2101.08641 [astro-ph.CO]}
  \BibitemShut {NoStop}%
\bibitem [{\citenamefont {Aghanim}\ \emph
  {et~al.}(2020{\natexlab{b}})\citenamefont {Aghanim} \emph
  {et~al.}}]{Aghanim:2018eyx}%
  \BibitemOpen
  \bibfield  {author} {\bibinfo {author} {\bibfnamefont {N.}~\bibnamefont
  {Aghanim}} \emph {et~al.} (\bibinfo {collaboration} {Planck}),\ }\href
  {\doibase 10.1051/0004-6361/201833910} {\bibfield  {journal} {\bibinfo
  {journal} {Astron. Astrophys.}\ }\textbf {\bibinfo {volume} {641}},\ \bibinfo
  {pages} {A6} (\bibinfo {year} {2020}{\natexlab{b}})},\ \Eprint
  {http://arxiv.org/abs/1807.06209} {arXiv:1807.06209 [astro-ph.CO]}
  \BibitemShut {NoStop}%
\bibitem [{\citenamefont {Alcaniz}\ \emph {et~al.}(2021)\citenamefont
  {Alcaniz}, \citenamefont {Bernal}, \citenamefont {Masiero},\ and\
  \citenamefont {Queiroz}}]{Alcaniz:2019kah}%
  \BibitemOpen
  \bibfield  {author} {\bibinfo {author} {\bibfnamefont {J.}~\bibnamefont
  {Alcaniz}}, \bibinfo {author} {\bibfnamefont {N.}~\bibnamefont {Bernal}},
  \bibinfo {author} {\bibfnamefont {A.}~\bibnamefont {Masiero}}, \ and\
  \bibinfo {author} {\bibfnamefont {F.~S.}\ \bibnamefont {Queiroz}},\ }\href
  {\doibase 10.1016/j.physletb.2020.136008} {\bibfield  {journal} {\bibinfo
  {journal} {Phys. Lett. B}\ }\textbf {\bibinfo {volume} {812}},\ \bibinfo
  {pages} {136008} (\bibinfo {year} {2021})},\ \Eprint
  {http://arxiv.org/abs/1912.05563} {arXiv:1912.05563 [astro-ph.CO]}
  \BibitemShut {NoStop}%
\bibitem [{\citenamefont {Kelso}\ \emph {et~al.}(2013)\citenamefont {Kelso},
  \citenamefont {Profumo},\ and\ \citenamefont {Queiroz}}]{Kelso:2013paa}%
  \BibitemOpen
  \bibfield  {author} {\bibinfo {author} {\bibfnamefont {C.}~\bibnamefont
  {Kelso}}, \bibinfo {author} {\bibfnamefont {S.}~\bibnamefont {Profumo}}, \
  and\ \bibinfo {author} {\bibfnamefont {F.~S.}\ \bibnamefont {Queiroz}},\
  }\href {\doibase 10.1103/PhysRevD.88.023511} {\bibfield  {journal} {\bibinfo
  {journal} {Phys. Rev. D}\ }\textbf {\bibinfo {volume} {88}},\ \bibinfo
  {pages} {023511} (\bibinfo {year} {2013})},\ \Eprint
  {http://arxiv.org/abs/1304.5243} {arXiv:1304.5243 [hep-ph]} \BibitemShut
  {NoStop}%
\bibitem [{\citenamefont {Tong}(2019)}]{tongLecturesCosmology}%
  \BibitemOpen
  \bibfield  {author} {\bibinfo {author} {\bibfnamefont {D.}~\bibnamefont
  {Tong}},\ }\href@noop {} {\bibfield  {journal} {\bibinfo  {journal}
  {Cambridge University}\ } (\bibinfo {year} {2019})}\BibitemShut {NoStop}%
\bibitem [{\citenamefont {Yang}\ \emph {et~al.}(2018)\citenamefont {Yang},
  \citenamefont {Pan}, \citenamefont {Di~Valentino}, \citenamefont {Nunes},
  \citenamefont {Vagnozzi},\ and\ \citenamefont {Mota}}]{Yang:2018euj}%
  \BibitemOpen
  \bibfield  {author} {\bibinfo {author} {\bibfnamefont {W.}~\bibnamefont
  {Yang}}, \bibinfo {author} {\bibfnamefont {S.}~\bibnamefont {Pan}}, \bibinfo
  {author} {\bibfnamefont {E.}~\bibnamefont {Di~Valentino}}, \bibinfo {author}
  {\bibfnamefont {R.~C.}\ \bibnamefont {Nunes}}, \bibinfo {author}
  {\bibfnamefont {S.}~\bibnamefont {Vagnozzi}}, \ and\ \bibinfo {author}
  {\bibfnamefont {D.~F.}\ \bibnamefont {Mota}},\ }\href {\doibase
  10.1088/1475-7516/2018/09/019} {\bibfield  {journal} {\bibinfo  {journal}
  {JCAP}\ }\textbf {\bibinfo {volume} {09}},\ \bibinfo {pages} {019} (\bibinfo
  {year} {2018})},\ \Eprint {http://arxiv.org/abs/1805.08252} {arXiv:1805.08252
  [astro-ph.CO]} \BibitemShut {NoStop}%
\bibitem [{\citenamefont {Aprile}\ \emph {et~al.}(2022)\citenamefont {Aprile}
  \emph {et~al.}}]{XENONCollaboration:2022kmb}%
  \BibitemOpen
  \bibfield  {author} {\bibinfo {author} {\bibfnamefont {E.}~\bibnamefont
  {Aprile}} \emph {et~al.} (\bibinfo {collaboration} {(XENON
  Collaboration)\textdagger{}\textdagger{}, XENON}),\ }\href {\doibase
  10.1103/PhysRevLett.129.161805} {\bibfield  {journal} {\bibinfo  {journal}
  {Phys. Rev. Lett.}\ }\textbf {\bibinfo {volume} {129}},\ \bibinfo {pages}
  {161805} (\bibinfo {year} {2022})},\ \Eprint
  {http://arxiv.org/abs/2207.11330} {arXiv:2207.11330 [hep-ex]} \BibitemShut
  {NoStop}%
\bibitem [{\citenamefont {Duarte}\ \emph {et~al.}(2022)\citenamefont {Duarte},
  \citenamefont {Lin}, \citenamefont {Lindner}, \citenamefont {Kozhuharov},
  \citenamefont {Kuleshov}, \citenamefont {de~Jesus}, \citenamefont {Queiroz},
  \citenamefont {Villamizar},\ and\ \citenamefont {Westfahl}}]{Duarte:2022feb}%
  \BibitemOpen
  \bibfield  {author} {\bibinfo {author} {\bibfnamefont {L.}~\bibnamefont
  {Duarte}}, \bibinfo {author} {\bibfnamefont {L.}~\bibnamefont {Lin}},
  \bibinfo {author} {\bibfnamefont {M.}~\bibnamefont {Lindner}}, \bibinfo
  {author} {\bibfnamefont {V.}~\bibnamefont {Kozhuharov}}, \bibinfo {author}
  {\bibfnamefont {S.~V.}\ \bibnamefont {Kuleshov}}, \bibinfo {author}
  {\bibfnamefont {A.~S.}\ \bibnamefont {de~Jesus}}, \bibinfo {author}
  {\bibfnamefont {F.~S.}\ \bibnamefont {Queiroz}}, \bibinfo {author}
  {\bibfnamefont {Y.}~\bibnamefont {Villamizar}}, \ and\ \bibinfo {author}
  {\bibfnamefont {H.}~\bibnamefont {Westfahl}},\ }\href@noop {} {\  (\bibinfo
  {year} {2022})},\ \Eprint {http://arxiv.org/abs/2206.05305} {arXiv:2206.05305
  [hep-ph]} \BibitemShut {NoStop}%
\bibitem [{\citenamefont {Feng}\ \emph {et~al.}(2003)\citenamefont {Feng},
  \citenamefont {Rajaraman},\ and\ \citenamefont {Takayama}}]{Feng:2003uy}%
  \BibitemOpen
  \bibfield  {author} {\bibinfo {author} {\bibfnamefont {J.~L.}\ \bibnamefont
  {Feng}}, \bibinfo {author} {\bibfnamefont {A.}~\bibnamefont {Rajaraman}}, \
  and\ \bibinfo {author} {\bibfnamefont {F.}~\bibnamefont {Takayama}},\ }\href
  {\doibase 10.1103/PhysRevD.68.063504} {\bibfield  {journal} {\bibinfo
  {journal} {Phys. Rev. D}\ }\textbf {\bibinfo {volume} {68}},\ \bibinfo
  {pages} {063504} (\bibinfo {year} {2003})},\ \Eprint
  {http://arxiv.org/abs/hep-ph/0306024} {arXiv:hep-ph/0306024} \BibitemShut
  {NoStop}%
\bibitem [{\citenamefont {Feng}\ \emph {et~al.}(2004)\citenamefont {Feng},
  \citenamefont {Su},\ and\ \citenamefont {Takayama}}]{Feng:2004zu}%
  \BibitemOpen
  \bibfield  {author} {\bibinfo {author} {\bibfnamefont {J.~L.}\ \bibnamefont
  {Feng}}, \bibinfo {author} {\bibfnamefont {S.-f.}\ \bibnamefont {Su}}, \ and\
  \bibinfo {author} {\bibfnamefont {F.}~\bibnamefont {Takayama}},\ }\href
  {\doibase 10.1103/PhysRevD.70.063514} {\bibfield  {journal} {\bibinfo
  {journal} {Phys. Rev. D}\ }\textbf {\bibinfo {volume} {70}},\ \bibinfo
  {pages} {063514} (\bibinfo {year} {2004})},\ \Eprint
  {http://arxiv.org/abs/hep-ph/0404198} {arXiv:hep-ph/0404198} \BibitemShut
  {NoStop}%
\bibitem [{\citenamefont {Alcaniz}\ \emph {et~al.}(2022)\citenamefont
  {Alcaniz}, \citenamefont {Neto}, \citenamefont {Queiroz}, \citenamefont
  {da~Silva},\ and\ \citenamefont {Silva}}]{Alcaniz:2022oow}%
  \BibitemOpen
  \bibfield  {author} {\bibinfo {author} {\bibfnamefont {J.~S.}\ \bibnamefont
  {Alcaniz}}, \bibinfo {author} {\bibfnamefont {J.~P.}\ \bibnamefont {Neto}},
  \bibinfo {author} {\bibfnamefont {F.~S.}\ \bibnamefont {Queiroz}}, \bibinfo
  {author} {\bibfnamefont {D.~R.}\ \bibnamefont {da~Silva}}, \ and\ \bibinfo
  {author} {\bibfnamefont {R.}~\bibnamefont {Silva}},\ }\href {\doibase
  10.1038/s41598-022-24608-5} {\bibfield  {journal} {\bibinfo  {journal} {Sci.
  Rep.}\ }\textbf {\bibinfo {volume} {12}},\ \bibinfo {pages} {20113} (\bibinfo
  {year} {2022})},\ \Eprint {http://arxiv.org/abs/2211.14345} {arXiv:2211.14345
  [astro-ph.CO]} \BibitemShut {NoStop}%
\bibitem [{\citenamefont {Klypin}\ \emph {et~al.}(1993)\citenamefont {Klypin},
  \citenamefont {Holtzman}, \citenamefont {Primack},\ and\ \citenamefont
  {Regos}}]{Klypin:1992sf}%
  \BibitemOpen
  \bibfield  {author} {\bibinfo {author} {\bibfnamefont {A.}~\bibnamefont
  {Klypin}}, \bibinfo {author} {\bibfnamefont {J.}~\bibnamefont {Holtzman}},
  \bibinfo {author} {\bibfnamefont {J.}~\bibnamefont {Primack}}, \ and\
  \bibinfo {author} {\bibfnamefont {E.}~\bibnamefont {Regos}},\ }\href
  {\doibase 10.1086/173210} {\bibfield  {journal} {\bibinfo  {journal}
  {Astrophys. J.}\ }\textbf {\bibinfo {volume} {416}},\ \bibinfo {pages} {1}
  (\bibinfo {year} {1993})},\ \Eprint {http://arxiv.org/abs/astro-ph/9305011}
  {arXiv:astro-ph/9305011} \BibitemShut {NoStop}%
\bibitem [{\citenamefont {Bose}\ \emph {et~al.}(2019)\citenamefont {Bose},
  \citenamefont {Vogelsberger}, \citenamefont {Zavala}, \citenamefont
  {Pfrommer}, \citenamefont {Cyr-Racine}, \citenamefont {Bohr},\ and\
  \citenamefont {Bringmann}}]{Bose:2018juc}%
  \BibitemOpen
  \bibfield  {author} {\bibinfo {author} {\bibfnamefont {S.}~\bibnamefont
  {Bose}}, \bibinfo {author} {\bibfnamefont {M.}~\bibnamefont {Vogelsberger}},
  \bibinfo {author} {\bibfnamefont {J.}~\bibnamefont {Zavala}}, \bibinfo
  {author} {\bibfnamefont {C.}~\bibnamefont {Pfrommer}}, \bibinfo {author}
  {\bibfnamefont {F.-Y.}\ \bibnamefont {Cyr-Racine}}, \bibinfo {author}
  {\bibfnamefont {S.}~\bibnamefont {Bohr}}, \ and\ \bibinfo {author}
  {\bibfnamefont {T.}~\bibnamefont {Bringmann}},\ }\href {\doibase
  10.1093/mnras/stz1276} {\bibfield  {journal} {\bibinfo  {journal} {Mon. Not.
  Roy. Astron. Soc.}\ }\textbf {\bibinfo {volume} {487}},\ \bibinfo {pages}
  {522} (\bibinfo {year} {2019})},\ \Eprint {http://arxiv.org/abs/1811.10630}
  {arXiv:1811.10630 [astro-ph.CO]} \BibitemShut {NoStop}%
\bibitem [{\citenamefont {Lopez-Corredoira}\ and\ \citenamefont
  {Marmet}(2022)}]{Lopez-Corredoira:2022dxt}%
  \BibitemOpen
  \bibfield  {author} {\bibinfo {author} {\bibfnamefont {M.}~\bibnamefont
  {Lopez-Corredoira}}\ and\ \bibinfo {author} {\bibfnamefont {L.}~\bibnamefont
  {Marmet}},\ }\href {\doibase 10.1142/S0218271822300142} {\bibfield  {journal}
  {\bibinfo  {journal} {Int. J. Mod. Phys. D}\ }\textbf {\bibinfo {volume}
  {31}},\ \bibinfo {pages} {2230014} (\bibinfo {year} {2022})},\ \Eprint
  {http://arxiv.org/abs/2202.12897} {arXiv:2202.12897 [astro-ph.CO]}
  \BibitemShut {NoStop}%
\bibitem [{\citenamefont {Lahav}\ and\ \citenamefont
  {Liddle}(2022)}]{Lahav:2022poa}%
  \BibitemOpen
  \bibfield  {author} {\bibinfo {author} {\bibfnamefont {O.}~\bibnamefont
  {Lahav}}\ and\ \bibinfo {author} {\bibfnamefont {A.~R.}\ \bibnamefont
  {Liddle}},\ }\href@noop {} {\  (\bibinfo {year} {2022})},\ \Eprint
  {http://arxiv.org/abs/2201.08666} {arXiv:2201.08666 [astro-ph.CO]}
  \BibitemShut {NoStop}%
\bibitem [{\citenamefont {Workman}\ \emph {et~al.}(2022)\citenamefont {Workman}
  \emph {et~al.}}]{ParticleDataGroup:2022pth}%
  \BibitemOpen
  \bibfield  {author} {\bibinfo {author} {\bibfnamefont {R.~L.}\ \bibnamefont
  {Workman}} \emph {et~al.} (\bibinfo {collaboration} {Particle Data Group}),\
  }\href {\doibase 10.1093/ptep/ptac097} {\bibfield  {journal} {\bibinfo
  {journal} {PTEP}\ }\textbf {\bibinfo {volume} {2022}},\ \bibinfo {pages}
  {083C01} (\bibinfo {year} {2022})}\BibitemShut {NoStop}%
\bibitem [{\citenamefont {Griffiths}(2020)}]{griffiths2020introduction}%
  \BibitemOpen
  \bibfield  {author} {\bibinfo {author} {\bibfnamefont {D.}~\bibnamefont
  {Griffiths}},\ }\href@noop {} {\emph {\bibinfo {title} {Introduction to
  elementary particles}}}\ (\bibinfo  {publisher} {John Wiley \& Sons},\
  \bibinfo {year} {2020})\BibitemShut {NoStop}%
\bibitem [{\citenamefont {Baring}\ \emph {et~al.}(2016)\citenamefont {Baring},
  \citenamefont {Ghosh}, \citenamefont {Queiroz},\ and\ \citenamefont
  {Sinha}}]{Baring:2015sza}%
  \BibitemOpen
  \bibfield  {author} {\bibinfo {author} {\bibfnamefont {M.~G.}\ \bibnamefont
  {Baring}}, \bibinfo {author} {\bibfnamefont {T.}~\bibnamefont {Ghosh}},
  \bibinfo {author} {\bibfnamefont {F.~S.}\ \bibnamefont {Queiroz}}, \ and\
  \bibinfo {author} {\bibfnamefont {K.}~\bibnamefont {Sinha}},\ }\href
  {\doibase 10.1103/PhysRevD.93.103009} {\bibfield  {journal} {\bibinfo
  {journal} {Phys. Rev. D}\ }\textbf {\bibinfo {volume} {93}},\ \bibinfo
  {pages} {103009} (\bibinfo {year} {2016})},\ \Eprint
  {http://arxiv.org/abs/1510.00389} {arXiv:1510.00389 [hep-ph]} \BibitemShut
  {NoStop}%
\bibitem [{\citenamefont {Mambrini}\ \emph {et~al.}(2016)\citenamefont
  {Mambrini}, \citenamefont {Profumo},\ and\ \citenamefont
  {Queiroz}}]{Mambrini:2015sia}%
  \BibitemOpen
  \bibfield  {author} {\bibinfo {author} {\bibfnamefont {Y.}~\bibnamefont
  {Mambrini}}, \bibinfo {author} {\bibfnamefont {S.}~\bibnamefont {Profumo}}, \
  and\ \bibinfo {author} {\bibfnamefont {F.~S.}\ \bibnamefont {Queiroz}},\
  }\href {\doibase 10.1016/j.physletb.2016.07.076} {\bibfield  {journal}
  {\bibinfo  {journal} {Phys. Lett. B}\ }\textbf {\bibinfo {volume} {760}},\
  \bibinfo {pages} {807} (\bibinfo {year} {2016})},\ \Eprint
  {http://arxiv.org/abs/1508.06635} {arXiv:1508.06635 [hep-ph]} \BibitemShut
  {NoStop}%
\bibitem [{\citenamefont {Wang}\ and\ \citenamefont
  {Yang}(2004)}]{Wang:2004ib}%
  \BibitemOpen
  \bibfield  {author} {\bibinfo {author} {\bibfnamefont {F.}~\bibnamefont
  {Wang}}\ and\ \bibinfo {author} {\bibfnamefont {J.~M.}\ \bibnamefont
  {Yang}},\ }\href {\doibase 10.1140/epjc/s2004-02029-6} {\bibfield  {journal}
  {\bibinfo  {journal} {Eur. Phys. J. C}\ }\textbf {\bibinfo {volume} {38}},\
  \bibinfo {pages} {129} (\bibinfo {year} {2004})},\ \Eprint
  {http://arxiv.org/abs/hep-ph/0405186} {arXiv:hep-ph/0405186} \BibitemShut
  {NoStop}%
\bibitem [{\citenamefont {Cembranos}\ \emph {et~al.}(2005)\citenamefont
  {Cembranos}, \citenamefont {Feng}, \citenamefont {Rajaraman},\ and\
  \citenamefont {Takayama}}]{Cembranos:2005us}%
  \BibitemOpen
  \bibfield  {author} {\bibinfo {author} {\bibfnamefont {J.~A.~R.}\
  \bibnamefont {Cembranos}}, \bibinfo {author} {\bibfnamefont {J.~L.}\
  \bibnamefont {Feng}}, \bibinfo {author} {\bibfnamefont {A.}~\bibnamefont
  {Rajaraman}}, \ and\ \bibinfo {author} {\bibfnamefont {F.}~\bibnamefont
  {Takayama}},\ }\href {\doibase 10.1103/PhysRevLett.95.181301} {\bibfield
  {journal} {\bibinfo  {journal} {Phys. Rev. Lett.}\ }\textbf {\bibinfo
  {volume} {95}},\ \bibinfo {pages} {181301} (\bibinfo {year} {2005})},\
  \Eprint {http://arxiv.org/abs/hep-ph/0507150} {arXiv:hep-ph/0507150}
  \BibitemShut {NoStop}%
\bibitem [{\citenamefont {Kelso}\ \emph {et~al.}(2014)\citenamefont {Kelso},
  \citenamefont {de~S.~Pires}, \citenamefont {Profumo}, \citenamefont
  {Queiroz},\ and\ \citenamefont {Rodrigues~da Silva}}]{Kelso:2013nwa}%
  \BibitemOpen
  \bibfield  {author} {\bibinfo {author} {\bibfnamefont {C.}~\bibnamefont
  {Kelso}}, \bibinfo {author} {\bibfnamefont {C.~A.}\ \bibnamefont
  {de~S.~Pires}}, \bibinfo {author} {\bibfnamefont {S.}~\bibnamefont
  {Profumo}}, \bibinfo {author} {\bibfnamefont {F.~S.}\ \bibnamefont
  {Queiroz}}, \ and\ \bibinfo {author} {\bibfnamefont {P.~S.}\ \bibnamefont
  {Rodrigues~da Silva}},\ }\href {\doibase 10.1140/epjc/s10052-014-2797-3}
  {\bibfield  {journal} {\bibinfo  {journal} {Eur. Phys. J. C}\ }\textbf
  {\bibinfo {volume} {74}},\ \bibinfo {pages} {2797} (\bibinfo {year}
  {2014})},\ \Eprint {http://arxiv.org/abs/1308.6630} {arXiv:1308.6630
  [hep-ph]} \BibitemShut {NoStop}%
\end{thebibliography}%

\end{document}